\documentclass[reprint,superscriptaddress,amsmath,amssymb,aps,floatfix,longbibliography]{revtex4-1}

\usepackage[pdftex]{graphicx}
\usepackage{bm}

\begin{document}

\title{Fluid-driven deformation of a soft granular material}

\author{Christopher W. MacMinn}
\email{christopher.macminn@eng.ox.ac.uk}
\affiliation{University of Oxford, Oxford, UK}
\affiliation{Yale University, New Haven, CT, USA}
\author{Eric R. Dufresne}
\affiliation{Yale University, New Haven, CT, USA}
\author{John S. Wettlaufer}
\affiliation{University of Oxford, Oxford, UK}
\affiliation{Yale University, New Haven, CT, USA}
\affiliation{Nordita, Royal Institute of Technology and Stockholm University, Stockholm, Sweden}

\date{\today}

\begin{abstract}
    Compressing a porous, fluid-filled material will drive the interstitial fluid out of the pore space, as when squeezing water out of a kitchen sponge. Inversely, injecting fluid into a porous material can deform the solid structure, as when fracturing a shale for natural gas recovery. These poromechanical interactions play an important role in geological and biological systems across a wide range of scales, from the propagation of magma through the Earth's mantle to the transport of fluid through living cells and tissues. The theory of poroelasticity has been largely successful in modeling poromechanical behavior in relatively simple systems, but this continuum theory is fundamentally limited by our understanding of the pore-scale interactions between the fluid and the solid, and these problems are notoriously difficult to study in a laboratory setting. Here, we present a high-resolution measurement of injection-driven poromechanical deformation in a system with granular microsctructure: We inject fluid into a dense, confined monolayer of soft particles and use particle tracking to reveal the dynamics of the multi-scale deformation field. We find that a continuum model based on poroelasticity theory captures certain macroscopic features of the deformation, but the particle-scale deformation field exhibits dramatic departures from smooth, continuum behavior. We observe particle-scale rearrangement and hysteresis, as well as petal-like mesoscale structures that are connected to material failure through spiral shear banding.
\end{abstract}

\maketitle

\section{Introduction}

Poromechanics couples the mechanical deformation of a porous solid with fluid flow through its internal structure~\cite{terzaghi-procsmfe-1936, biot-jap-1941, atkin-imajapplmath-1976, kenyon-archrationmechanalysis-1976b, coussy-wiley-2004}. In biophysics, poromechanics plays an important role in the growth and deformation of cells and tissues~(\textit{e.g.}, \cite{yang-jbiomech-1991, lai-jbiomecheng-1991, cowin-jbiomech-1999, charras-nature-2005, moeendarbary-natmaterials-2013}), and it is the dominant mechanism underlying plant motion~(\textit{e.g.}, \cite{dumais-annrevfluidmech-2012}). In both pure and applied geophysics, poromechanics has been studied intensely in the context of subsurface pressurization during fluid injection, such as in geothermal energy extraction or carbon dioxide sequestration~(\textit{e.g.}, \cite{szulczewski-pnas-2012, nrc-nationalacademiespress-2013, verdon-pnas-2013, jha-wrr-2014}), and particularly in the context of hydraulic fracture for enhanced oil or gas recovery (\textit{e.g.}, \cite{hubbert-transaime-1957, detournay-intjrmms-1988, yarushina-geophysjint-2013}).

Poromechanical deformations are poro\textit{elastic} when they are controlled by the reversible storage and release of elastic energy. The classical theory of poroelasticity couples linear elasticity with Darcy's law for fluid flow through a porous medium, and the hallmark of these systems is diffusive propagation and dissipation of fluid pressure with characteristic time scale $T_\mathrm{pe}=\mu{}L^2/(\mathcal{K}k)$, where $\mu$ is the viscosity of the fluid, $L$ is a characteristic length scale, and $\mathcal{K}$ and $k$ are the elastic modulus and permeability of the solid skeleton. This approach is valid for small deformations, but many real systems feature large deformations, small-scale microstructure, or physical mechanisms such as damage, growth, or swelling, that lead to a strongly nonlinear coupling between the pore structure and the fluid flow~\cite{wang-princeton-2000, coussy-wiley-2004}.

Many poroelastic deformations of practical interest are driven by fluid injection. Injection-driven deformations involve radial dilation (outward expansion), which is particularly interesting and challenging because it leads to a nontrivial state of stress and strain~(\textit{e.g.}, \cite{yu-geotechnique-1991, alsiny-geotechnique-1992, hutchens-softmatter-2014, coulais-prl-2014}). Indeed, fluid injection into granular materials can lead to spectacular damage patterns when the injection pressure exceeds the inter-particle friction or the external confining stress~(\textit{e.g.}, \cite{johnsen-pre-2006, cheng-natphys-2008, sandnes-natcomms-2011, huang-prl-2012}). However, the deformation in these examples is almost completely irreversible because the solid skeleton is stiff and the fluid pressure is low, so the majority of the input energy is dissipated through frictional sliding and rearrangement, and almost none is stored elastically~\cite{holtzman-prl-2012}. Fluid injection can only drive significant storage of elastic energy when the fluid pressure becomes comparable to the stiffness of the solid skeleton. 

True poroelastic deformation requires either much larger pressures or much softer materials. As a result, it has proven difficult to study in a laboratory setting. Experiments with rocks and sands have been limited to postmortem inspection after high-pressure injection~(\textit{e.g.}, \cite{bohloli-jpetscieng-2006, germanovich-spe-2012}), providing useful insight into the failure of realistic geomaterials, but at a very coarse level in time and space. This limitation has been avoided in a one-dimensional geometry with soft, open-cell polymer foams (kitchen sponges; \textit{e.g.}, \cite{beavers-japplmech-1975, parker-japplmech-1987, sobac-mecind-2011}). However, these materials have proven to be experimentally challenging for a variety of reasons, with unsatisfying comparison between experiment and theory.

Here, we study the poromechanical deformation of a system with granular microstructure by injecting fluid into a confined monolayer of spherical particles. By using particles that are extremely soft, we construct a system that exhibits striking poroelastic phenomena at relatively low working pressures. High-resolution imaging and particle-tracking provide experimental access to the full, multi-scale deformation field. We show that the smooth, quasi-reversible macroscopic deformation can be captured in part by a minimal continuum model, despite the presence of complex shear banding and structural rearrangement.

\section{Fluid injection into \\ a monolayer of soft particles}

\begin{figure*}
    \centering
    \includegraphics[width=17.2cm]{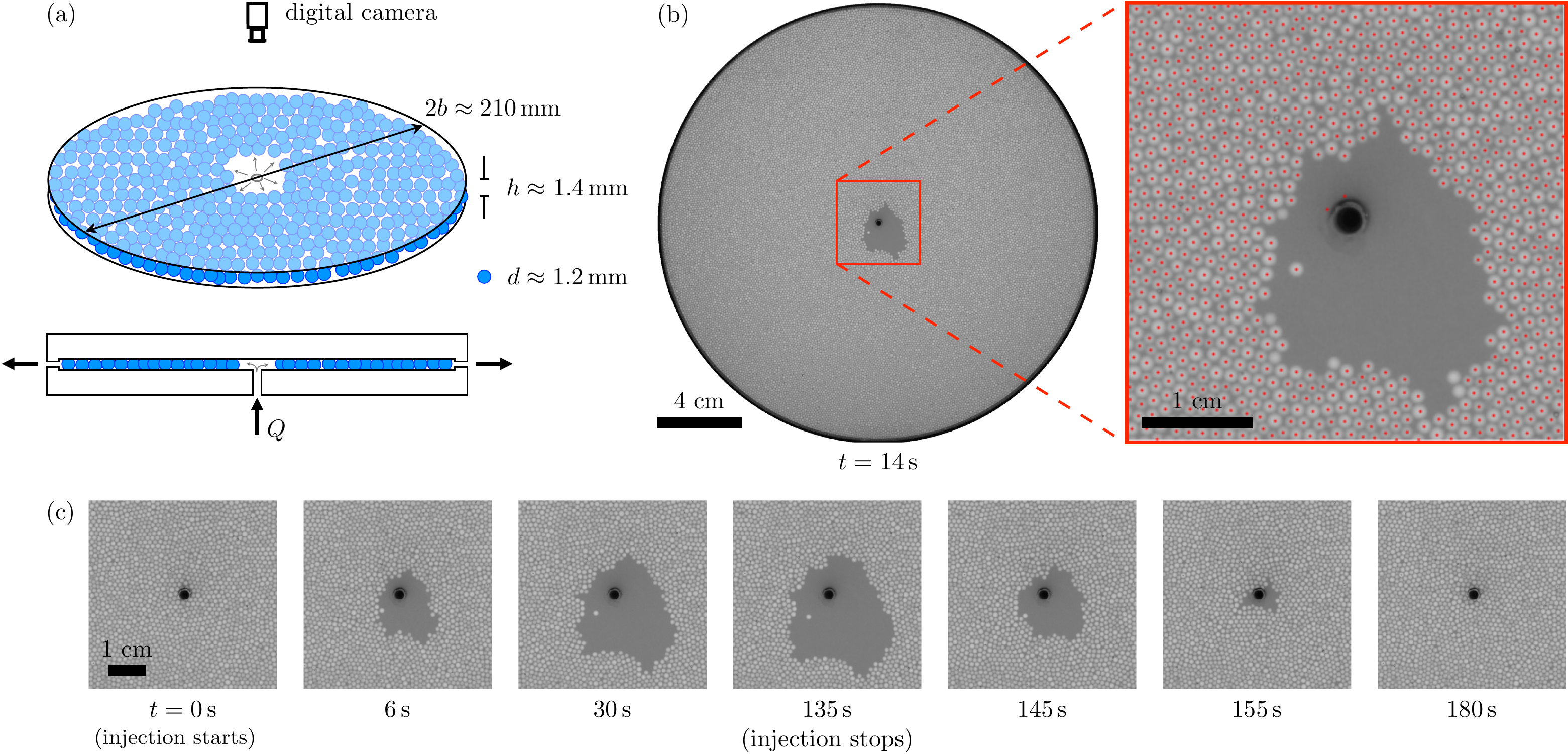}
    \caption{We inject fluid into a monolayer of soft particles and measure the resulting poromechanical deformation. (a)~The particles are confined between two rigid glass discs of radius $b\approx$105~mm, and we inject fluid into the center of the packing at a steady volume rate $Q$. The fluid flows radially outward through the packing and exits along the rim through a permeable spacer. The pressure gradient due to fluid flow through the packing drives the particles radially outward, opening a cavity in the center that relaxes and closes after injection stops. (b)~We image the experiment from above with a digital still camera, measuring the deformation field at high resolution by identifying and tracking the individual particles (see Appendix~A). This snapshot is during injection at $Q=$24~mL/min.  (c)~Sequence of snapshots from the same experiment showing cavity opening and relaxation. The black circle in the center is the injection port in the bottom disc, which has a diameter of about 2.5~mm. \label{fig:experiments} }
\end{figure*}

We pack a single layer of about 25,000 spherical, polyacrylamide hydrogel particles between two glass discs and we saturate the packing with the working fluid, a mixture of water and glycerol~(Figure~1a; see Appendix~A). The discs are separated by a permeable spacer that confines the particles, but allows fluid to leave freely around the edge. The particles are soft (having a Young modulus of $\sim$20~kPa), nearly incompressible (having a Poisson ratio of $\sim$$1/2$), Hertzian (exhibiting Hertz-like contact mechanics), elastic (allowing order-one elastic strains), non-cohesive, and very slippery (having low friction at particle-particle and particle-wall contacts)~\cite{mukhopadhyay-pre-2011, brodu-arxiv-2014}. The particles have mean diameter $d\approx$1.2~mm with about 10\% polydispersity~(Figure~5). The packing has an apparent area void fraction of $\sim$0.14 when viewed from above, and an actual volume void fraction of $\sim$0.51. The former is denser than random close packing in 2D ($\sim$0.18) due in part to the softness of the particles, as often occurs in fluid emulsions, and also in part to their polydispersity.

To perform an experiment, we inject more of the same working fluid into the cell at a constant volume rate $Q$. This fluid enters the cell via an injection port in the center of the lower disc, flows radially outward through the packing, and exits through the spacer at the outer edge. The resulting fluid pressure gradient within the porous layer (large pressure at the center dropping to atmospheric at the edge) deforms the packing, driving the particles radially outward and opening a cavity in the center~(Figure~1b,c; Video~S1). This coupling of fluid pressure and solid deformation is the core idea behind poromechanics. The deformation eventually reaches a steady state (here, after $\sim$100~s) in which the gradient of elastic stress in the solid skeleton balances the gradient of pressure in the fluid. We then stop injecting, at which point the elastic stress relaxes as the pressure gradient dissipates and the cavity closes. The relaxation of the packing highlights the macroscopically \textit{elastic} nature of the system, demonstrating that the packing stores elastic energy during the injection phase and releases it during the relaxation phase. We repeat this injection-relaxation cycle several times in the same packing. We image the deformation and subsequent relaxation of the packing with a digital still camera, detecting the particle positions in each image to within about $0.01d$ and then tracking the particles from image to image~(Figure~1b; see Appendix~A).

\section{Multi-scale deformation field}

One striking aspect of the deformation is the cavity that opens and then closes in the center of the packing. Despite the irregular shape of the cavity (Figure~2b), we find that the macroscopic dynamics of its expansion and collapse are smooth and relatively reproducible across repeated injected-relaxation cycles~(Figures~2a and~7). In contrast, the shape of the cavity varies from cycle to cycle~(Figure~2b; Video~S2). The size of the cavity increases with injection rate roughly in accordance with the prediction of a minimal continuum model (described in more detail below), but repeating an experiment at a given injection rate after ``resetting'' the packing by completely rearranging the particles leads to large variability~(Figure~2c). This implies that the macroscopic properties of the packing are a strong function of particle arrangement.

\begin{figure*}
   \centering
   \includegraphics[width=17.2cm]{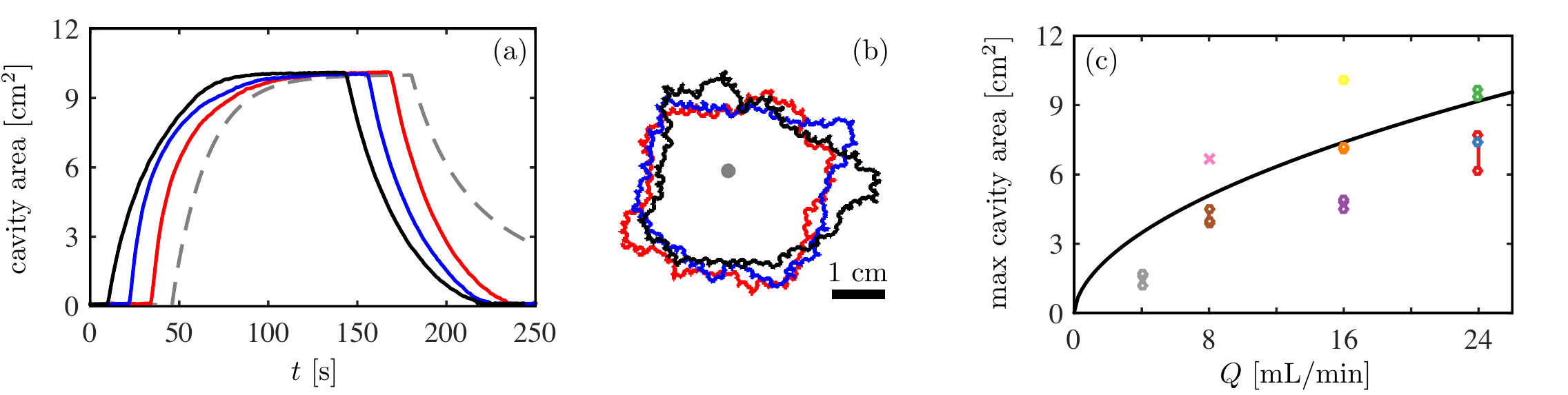}
   \caption{Fluid injection into the center of the packing drives outward compaction, opening a cavity. When injection stops, the packing relaxes and the cavity closes. Here, (a--b)~we inject at a steady rate of $Q=16$~mL/min for about 135~s and then stop, allowing the packing to relax. We repeat this injection-relaxation cycle two more times (black, then blue, then red). (a)~Comparing the area-vs-time curves from the three cycles (offset by 12~s for comparison; \textit{cf.} Figure~7) shows that the process is macroscopically smooth, with similar dynamics and maximum area in each cycle. The first cycle (black) is somewhat different than subsequent cycles due to heterogeneities in the initial particle distribution. We also show the prediction of a continuum model using best-fit material properties~(dashed gray curve); we discuss the model in \S{}V. (b)~The steady-state cavity shape is neither smooth nor repeatable, indicating the presence of irreversible micromechanics. The cavity does not open symmetrically about the injection port (gray circle with diameter $\sim$2.5~mm). The scalloped edges of the cavity profiles are particle-scale roughness~($\sim$1~mm). (c)~We repeat this experiment at different injection rates, showing here the maximum (\textit{i.e.}, steady-state) cavity area against injection rate $Q$. Each group (color) of circles indicates a series of least two cycles after ``resetting'' the packing by removing the particles, cleaning the apparatus, and replacing the same particles (the cross is a single cycle). The variability between cycles is relatively small except in one case (red), indicating that cycle-to-cycle irreversibilities have a weak impact on the macroscopic mechanics. The variability between series at the same injection rate is much larger, indicating that the particle arrangement has a strong impact on the macroscopic mechanics. All packings have initial porosity between 0.506 and 0.515, and we do not observe any clear correlation between initial porosity and cavity size in this range. The black curve is the prediction of the continuum model using the same mechanical properties for all points. \label{fig:cavity} }
\end{figure*}

We measure the internal deformation of the packing via particle tracking, which provides a direct measure of the displacement field. For this purpose, we define a rectangular coordinate system centered at the injection port, where $(x_i,y_i)$ is the position of particle $i$ at time $t$ and $(X_i,Y_i)$ is its initial position. The displacement of particle $i$ is then $\boldsymbol{u}_i=(x_i-X_i,y_i-Y_i)$, with magnitude $u_i(t)=\sqrt{(x_i-X_i)^2+(y_i-Y_i)^2}$ and radial component $u_{r,i}(t)=\sqrt{x_i^2+y_i^2}-\sqrt{X_i^2+Y_i^2}$. The deformation is primarily radial because of the axisymmetric boundary conditions, so we focus on $u_r$. The difference $u-|u_r|$ is a measure of the non-radial component of the displacement, which we find to be a few percent or less of the radial component~(see Appendix~A).

We find that the radial displacement is large near the cavity and fades to zero at the rigid edge, with a petal-like mesoscale structure (Figure~3a; Video~S3). Similar petal-like features have been observed in simulations of fluid injection into an initially dry packing of frictional particles~\cite{zhang-arma-2011}, indicating that these structures are not an artifact of our low-friction system. Additionally, similar but much more regular features have been observed in quasi-static ``grain injection'' experiments~\cite{pinto-prl-2007}, where they were identified with preferential directions in the far-field crystal structure. However, our packings are isotropic due to the polydispersity of the particles.

Each petal represents a group of particles that move radially outward further than their neighbors, implying that the edges of each petal are bands of localized shear failure. We confirm this by calculating the local strain field~(\cite{falk-pre-1998}; see Appendix~A), revealing a network of spiral shear bands that span the entire system~(Figure~3b). Shear bands following logarithmic spirals are a well-known feature of failure in radial dilation and hydraulic fracture~\cite{alsiny-geotechnique-1992, bohloli-jpetscieng-2006}. We see spirals with a pitch of roughly of $\pi/4$, which implies that the packing has a very low shear strength~(\textit{i.e.}, a very low friction angle). Correlations between shear strain and positive volumetric strain are evidence of shear dilation, a well-known feature of deformation in granular materials~(Figure~3b,c).



\begin{figure*}
   \centering
   \includegraphics[width=17.2cm]{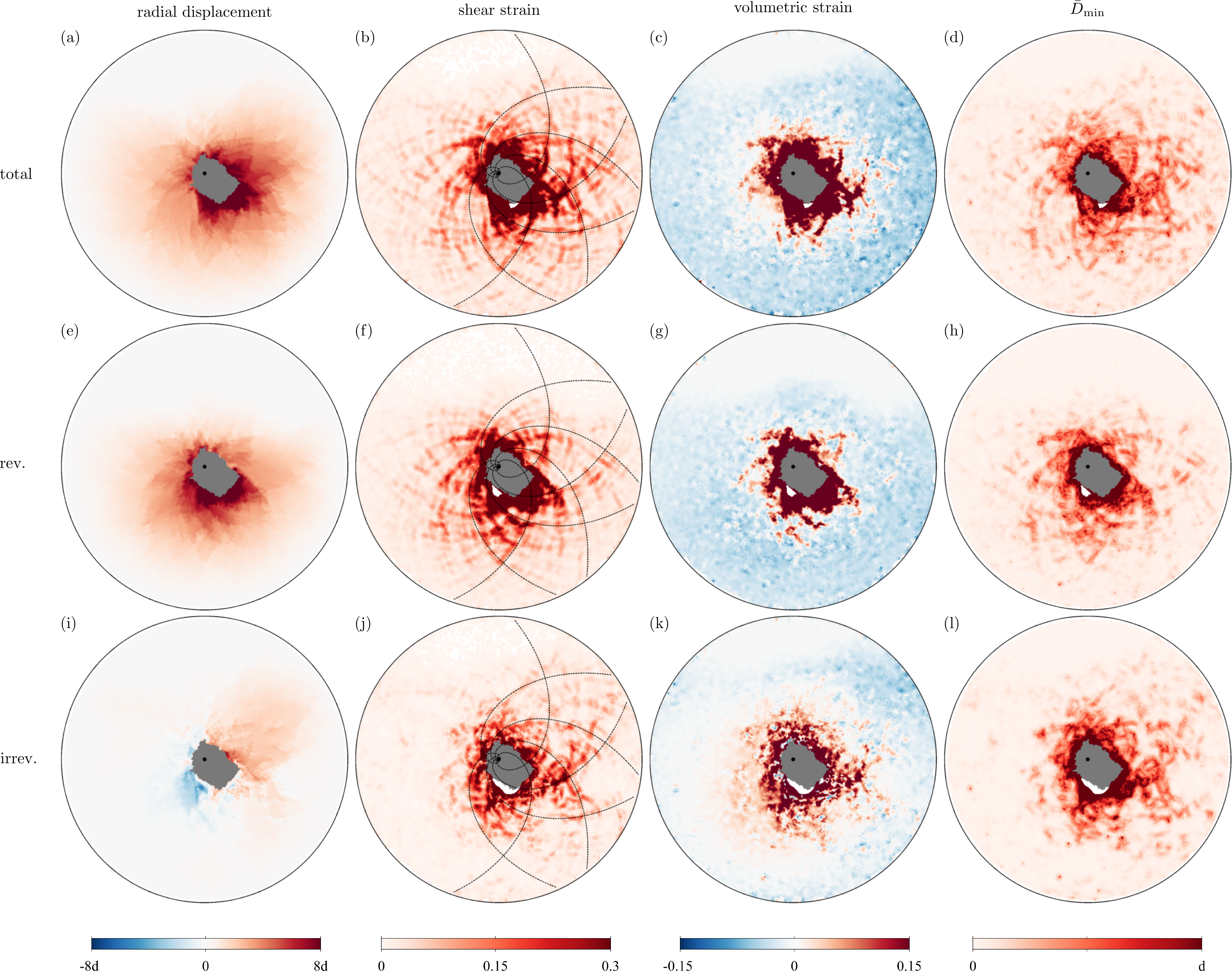}
   \caption{The displacement field is characterized by a petal-like mesoscale structure that corresponds to spiral bands in the shear strain field. Here we show (a)~the radial displacement, which reveals detailed mesoscale structure reminiscent of flower petals. We show (b)~the magnitude of the shear strain with several logarithmic spirals for comparison (dashed black). The spirals have a pitch of $\pi/4$, which is consistent with the shear failure of a material with negligible shear strength. We also show (c)~the volumetric strain. The volumetric strain shows expansion in the inner region and compression in the outer region, and is relatively axisymmetric. Lastly, we include (d)~the typical non-affine displacement $\bar{\boldsymbol{D}}_\mathrm{min}$ (see Appendix~A). The non-affine displacement is much smaller than the radial displacement except near the cavity, where the kinematics become strongly non-affine. The strain field is not necessarily a good measure of the deformation in areas with large amounts of non-affinity (near the cavity). We plot these quantities from an Eulerian perspective (against current radial position $r$). We decompose the deformation field into (e--h)~reversible and (i--l)~irreversible components. For comparison, we project all fields onto the deformed configuration at steady-state. The reversible component is larger in magnitude than the irreversible component, and contains most of the mesoscale structure. The irreversible component contains much of the non-affine displacement. \label{fig:strains} }
\end{figure*}

\section{Elasticity, plasticity, \\ (ir)reversibility, and dissipation}

Macroscopically, elastic deformations involve the reversible storage and release of strain energy, whereas plastic deformations are dissipative and irreversible. In crystalline solids, there is a clear distinction between elasticity and plasticity at the particle scale: Elastic deformations involve stretching or compressing bonds between atoms or molecules, whereas plastic deformations involve breaking and/or rearranging bonds. This distinction is less clear in amorphous or granular materials, where deformations often involve a combination of reversible and irreversible rearrangement events~\cite{lundberg-pre-2008, keim-softmatter-2013, keim-prl-2014}.

Here, the fact that the cavity closes completely upon relaxation implies that the deformation is macroscopically reversible. However, the hysteresis in cavity shape is evidence of particle-scale irreversibility, and the shear bands are evidence of plastic failure. To investigate the apparent contradiction of strongly irreversible micromechanics coexisting with smooth, quasi-reversible macroscopic mechanics, we decompose the deformation field into reversible and irreversible components. We calculate these by considering the transformation between three configurations: the initial (\textit{before} the deformation), the deformed (at steady-state), and the final (relaxed, \textit{after} the deformation). The total strain $\boldsymbol{E}$ is that which transforms the initial configuration into the deformed one. The irreversible strain $\boldsymbol{E}_\mathrm{irr}$ is that which remains after the deformation relaxes (\textit{i.e.}, the residual strain); this transforms the initial configuration to the final relaxed one. The reversible strain $\boldsymbol{E}_\mathrm{rev}$ is that which dissipates as the deformation relaxes; this would transform the final relaxed configuration back to the deformed one. For infinitesimal deformations, these three strains are related by superposition, $\boldsymbol{E}=\boldsymbol{E}_\mathrm{rev}+\boldsymbol{E}_\mathrm{irr}$; we calculate them independently since the deformation is large. Since we calculate strain as the locally affine best fit to the actual deformation field, we also calculate the root-mean-square difference between the affine field and the actual deformation field, $\bar{\boldsymbol{D}}_\mathrm{min}$ (\cite{falk-pre-1998}; see Appendix~A). This is a measure of the typical non-affine displacement, which is indicative of the amount of particle-level rearrangement.

Comparing the reversible and irreversible components of the strain field (Figure~3f,g vs. j,k), we find that the inner region is dominated by a combination of reversible and irreversible volumetric expansion (positive volumetric strain) and shear. Volumetric expansion indicates that particles have traveled away from their neighbors, which is expected near the cavity since the particles move radially outward by several diameters. Since the packing cannot support tension, this leads to local collapse or ``unjamming'' of the packing structure, which leads to large amounts of both reversible and irreversible rearrangement (Figure~3d,h,l). In contrast, the outer region is dominated by relatively smooth, axisymmetric, reversible volumetric compression (negative volumetric strain). The displacement and shear strain are much smaller, and there is much less rearrangement. The spiral shear bands span the entire system and, surprisingly, are primarily reversible~(Figure~3b,f,g).

All rearrangements play a strong role in the dynamics since reconfiguration of the packing takes time and dissipates energy. Macroscopically, dissipative deformations that are reversible are known as ``viscoelastic'', whereas those that are irreversible are known as ``viscoplastic''. Unlike elasticity, which is quasi-static, these viscous processes are rate dependent.

\section{Poroviscoelastic \\ continuum model}

The steady-state deformation is set by the balance between the gradient in fluid pressure within the packing and the roughly axisymmetric elastic compression of the outer region of the packing. Motivated by this, we next derive a minimal axisymmetric model for this system based on the theory of poroelasticity~\cite{coussy-wiley-2004}. Our model is intended to capture four main features of this system: (1)~conservation of volume, (2)~poromechanical coupling between pressure gradients in the fluid and stress gradients in the solid skeleton, (3)~elastic energy storage in the solid skeleton, and (4)~viscous dissipation due to reversible and irreversible rearrangements. We do not attempt to capture the evolution of effective material properties due to irreversible rearrangements (\textit{c.f.}, Figure~\ref{fig:cavity}c). We emphasize that we are not attempting to develop a general model for deforming granular materials, but rather a minimal one that captures the leading order behavior of our poroelastic system. We outline the core assumptions of the model here and present a detailed derivation in Appendix~B.

We assume that the packing is homogeneous, and that the flow and deformation fields are axisymmetric. We also assume that the fluid and the solid are individually incompressible which, for the solid, implies that the beads can rearrange and deform without changing volume. This is justified here because the working pressure is low relative to the bulk moduli of the fluid and the particles~($\sim$5~kPa \textit{vs.} $\sim$2~GPa). This allows for a simple but exact kinematic relationship between the volumetric strain and the local porosity (fluid or void fraction)~$\phi_f(r,t)$~(Equation~\ref{eq:phi_to_u}). We assume that the porosity is initially uniform and equal to $\phi_f(r,0)=\phi_{f,0}$.

We assume that the elastic stress in the solid skeleton is isotropic, meaning that the skeleton stores elastic energy due to volumetric compression but not due to shear. This is justified by the fact that similar packings are known to have an anomalously low ratio of shear to bulk modulus due to the extreme softness and slipperiness of the contacts~\cite{lietor-santos-pre-2011}. This allows us to link the stress directly to the volumetric strain, and therefore to $\phi_f$. This is useful because although the displacements are large, the changes in porosity are small. We therefore take the elastic component of the stress to be a power law in the change in porosity with exponent $3/2$, as appropriate for a granular material consisting of Hertzian particles~\cite{ohern-pre-2003}, with an effective (drained) bulk modulus $\mathcal{K}$.

The shear strain field indicates that the skeleton experiences shear failure near the cavity and along the spiral shear bands. The shear failure of granular materials is typically modelled with a frictional (Mohr-Coulomb) failure criterion, which states that the material will yield (fail plastically) when the shear stress anywhere exceeds some fixed fraction of the local normal stress. After yielding, the structure of the material rearranges according to a suitable (visco)plastic flow law. This transition to plastic flow (unjamming) has been studied extensively from a variety of perspectives~(\textit{e.g.}, \cite{falk-pre-1998, ohern-pre-2003, nordstrom-prl-2010, boyer-prl-2011}).

Although the dynamics of unjamming can be extremely important in many systems, we do not treat this behavior here because our system is \textit{compaction-controlled}. That is, the rate of shear strain near the cavity is fundamentally limited by the rate of volumetric strain in the outer region since the cavity can only expand as fast the outer region compresses. This is consistent with the fact that we expect the effective viscosity of rearrangement in the compacting outer region to be much larger than that in the unjamming inner region~\cite{boyer-prl-2011}. In an expansion-controlled system where the outer region is not yet jammed, we would expect the cavity shape to exhibit regular, sharp, triangular cracks~\cite{bandi-epl-2011}. In contrast, we see irregular and relatively smooth cavity shapes.

We model viscous dissipation due to rearrangement in the outer region in a very simplistic way by assuming that this contributes a transient component to the volumetric stress that is linear in the rate of change of porosity with an effective viscosity $\eta$. This linear, Kelvin-Voigt-like representation is often used in viscoelasticity, but here it is intended to capture both reversible and irreversible rearrangements. The linear viscous term embodies the rate-dependence of these processes, introducing a characteristic time scale for viscous rearrangement, $T_\mathrm{vr}=\eta/\mathcal{K}$. Although the viscosity should itself be a function of the local volumetric stress~\cite{boyer-prl-2011}, we ignore this for simplicity.

Finally, we assume that fluid flows relative to the solid skeleton according to Darcy's law with a constant permeability $k$. The assumption of constant permeability is justified by the fact that the changes in porosity are relatively small, at most a few percent.

These assumptions lead to a nonlinear conservation law for the local porosity as a function of time:
\begin{equation}\label{eq:model_nd}
    \frac{\partial{\tilde{\phi}_f}}{\partial{\tilde{t}}}+ \frac{1}{\tilde{r}}\frac{\partial}{\partial{\tilde{r}}}\left[\alpha\tilde{\phi}_f -\tilde{r}(1-\tilde{\phi}_f)\frac{\partial{\tilde{\sigma}^\prime}}{\partial{\tilde{r}}}\right]=0, 
\end{equation}
where $\tilde{\phi}_f=(\phi_f-\phi_{f,0})/(1-\phi_{f,0})$ is the normalized change in porosity, which is a measure of the Eulerian volumetric strain; $\tilde{t}$ and $\tilde{r}$ are time and radial position scaled by the poroelastic time scale $T_\mathrm{pe}=\mu{}b^2/(\mathcal{K}k)$ and the outer radius~$b$, respectively; and
\begin{equation}\label{eq:sigmaprime_nd}
    \tilde{\sigma}^\prime=\tilde{\phi}_f\,|\tilde{\phi}_f|^{1/2} +\beta\frac{\partial{\tilde{\phi}_f}}{\partial{\tilde{t}}}
\end{equation}
is the stress in the solid skeleton (the effective stress) scaled by the bulk modulus $\mathcal{K}$. The model has two dimensionless parameters, $\alpha=\mu{}Q/(2\pi{}h\mathcal{K}k)$ and $\beta=\eta{}k/(\mu{}b^2)$. The former compares the pressure gradient in the fluid with the stiffness of the solid skeleton, while the latter compares the viscous time scale to the poroelastic one. Because the mechanical properties of the packing are difficult to measure and strongly dependent on the particle arrangement, we use $\alpha$ and $\beta$ as fitting parameters. The former influences the rate of deformation and sets the steady-state; the latter influences only the rate.

We solve Eq.~\eqref{eq:model_nd} subject to two boundary conditions, which are that the cavity wall is a free surface where the stress in the solid skeleton vanishes, and that the outer edge is a rigid boundary where the solid is stationary. We also solve simultaneously an evolution equation for the position of the cavity wall. This model is a radial version of those that have been developed for the rectilinear deformation of kitchen sponges~\cite{sobac-mecind-2011}. A similar radial model was developed in the context of blood-vessel pressurization~\cite{barry-jaustralmathsocb-1993}, but our model is kinematically exact for finite axisymmetric deformations~(see Appendix~B) and we incorporate a nonlinear elastic stiffness as well as viscous dissipation.

\begin{figure*}[t]
   \centering
   \includegraphics[width=17.2cm]{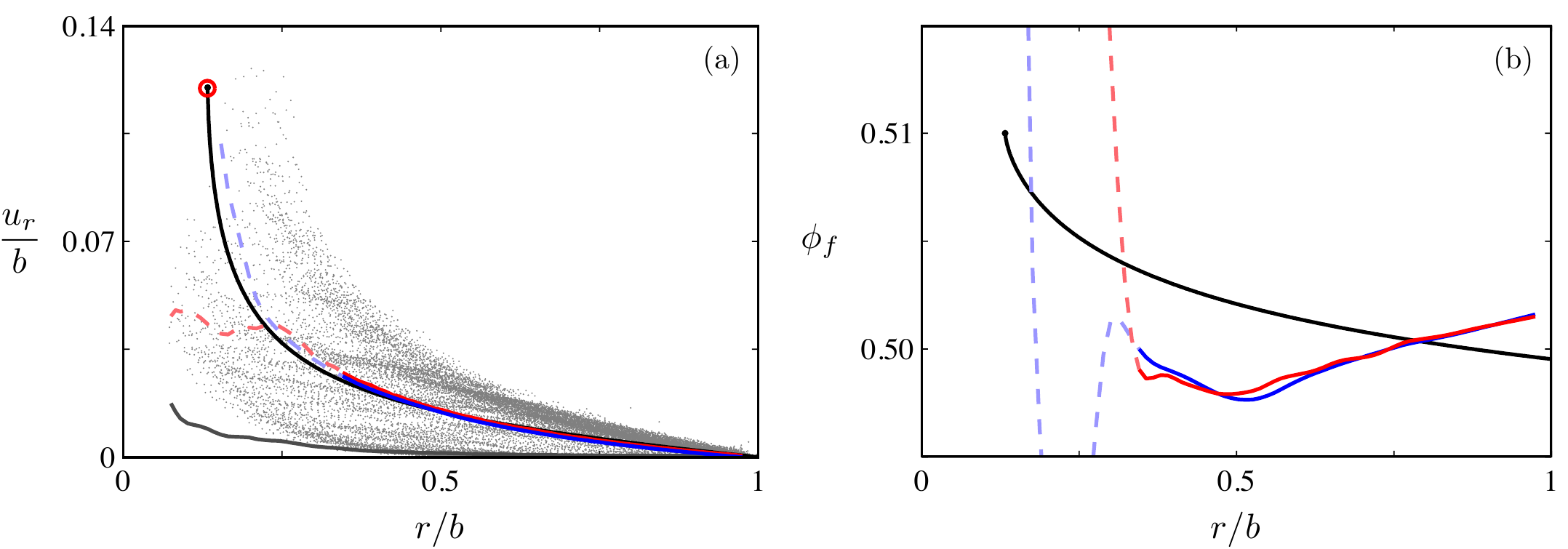}
   \caption{Despite the large and structured variability in the particle displacements, the continuum model agrees well with the azimuthally averaged displacement profile after choosing $\alpha$ to match the cavity radius. Here we show (a)~the steady-state radial displacement of the particles~(gray dots), the radius of the cavity~(red circle), and two different measures of the azimuthally averaged radial displacement (red curve: direct azimuthal average; blue curve: integrated volumetric strain). We also show the azimuthally averaged non-affine displacement (solid gray curve) and the displacement field predicted by the continuum model~(solid black curve). The displacement becomes strongly non-affine near the cavity, so we choose a threshold to the left of which affine quantities are poor measures of the deformation: The dashed portions of the red and blue curves are where the non-affine displacement accounts for more than 10\% of the total. We calculate (b)~two corresponding measures of the azimuthally averaged porosity by differentiating the averaged displacement according to Eq.~\eqref{eq:phi_to_u} and smoothing the result~(red and blue curves). This is a poor measure of porosity near the cavity because it assumes a smooth, axisymmetric displacement field. The model (solid black curve) does not capture the porosity field very well. We plot these quantities from an Eulerian perspective. \label{fig:ur} }
\end{figure*}

In steady state, the model can be simplified to an integral coupled with a nonlinear algebraic equation for the cavity radius; this is straightforward to evaluate numerically. Despite the simplicity of the model and the large grain-scale variability in the experiments, we find that, after choosing $\alpha$ to match the cavity area, the steady-state displacement field predicted by the model agrees surprisingly well with the azimuthal average from the experiments~(Figure~4a). We also calculate an azimuthally averaged porosity field from the displacement field via Equation~\eqref{eq:phi_to_u}. The model does not agree very well with this~(Figure~4b), although it does capture certain aspects: The initial porosity in the experiment is $\phi_{f,0}\approx{}0.51$, and both the model and the experiment show a concave-up trend from a value near 0.51 near the cavity down to a value near 0.50 at the wall. The measured porosity field is non-monotonic with a minimum value at $r/b\approx{}0.5$, whereas the model is monotonic with its minimum at $r/b=1$, but the minimum values are similar.

The model is able to capture the displacement field relatively well because the displacement field is an integral measure of the deformation, dominated by the mean features and relatively insensitive to the local details. In contrast, the porosity field is a direct measure of the local strain, strongly reflecting the local details of the deformation. In addition, use of Equation~\eqref{eq:phi_to_u} assumes smoothness, axisymmetry, and an initially uniform porosity field, but the experiment does not necessarily observe any these except in an averaged sense.

A more elaborate model might take advantage of plastic-failure theory from soil mechanics~\cite{yu-geotechnique-1991}, the theory of shear-transformation zones in amorphous solids~\cite{falk-pre-1998, falk-annrevcondmattphys-2011}, or constitutive laws from the rheology of suspensions~\cite{boyer-prl-2011}, although any of these approaches would lead to a substantial increase in complexity and several additional constitutive parameters.

Solving the time-dependent model numerically, we find that it captures the dynamics of cavity expansion after choosing $\beta$ accordingly~(Figure~2a). The fit ($\beta\approx5$) yields a poroelastic time scale of $\sim$1.5~s and a viscous one of $\sim$5~s, implying that viscous rearrangement controls the overall rate of cavity expansion. Given the complex nature of this process, it is surprising that our linear Kelvin-Voigt model captures the dynamics of cavity expansion as well as it does. The model does not capture relaxation very well, relaxing much more slowly than the experiment. This is not surprising since the amount of viscous dissipation due to rearrangement is likely different during relaxation than during cavity expansion. Hysteresis in effective material properties is common in plasticity, soil mechanics, and granular materials, but our minimal model does not include this.


\section{Conclusions}

Fluid injection into a soft granular material drives deformation that is macroscopically poroelastic, despite rich micromechanical complexity. We found that the deformation in the inner region (near the injection port) is dominated by irreversible structural plasticity that leads to strong variations in the cavity shape, whereas the outer region deforms smoothly and reversibly. The latter ultimately supports the radially outward loading and controls the macroscopic mechanics of the steady state, such that the leading order features of the deformation can be captured relatively well with an axisymmetric continuum model. We expect this coexistence of microscopic irreversibility with macroscopic reversibility to be a strong feature of elastic dilational deformation in any system with a low ratio of shear strength to bulk stiffness. Many intriguing problems remain, such as connecting changes in grain-scale structure with the evolution of macroscopic properties, examining more sophisticated material models based on plasticity theory, and exploring the dynamics of fluid-driven shear failure.

More broadly, this system is a promising platform for high-resolution measurement of the dynamics of poromechanical deformation, and it has several additional features that we did not take advantage of here. For example, polyacrylamide hydrogel is closely index-matched with water, making it extremely well-suited to visualization in three-dimensional systems~\cite{mukhopadhyay-pre-2011, dijksman-rsi-2012, byron-expfluids-2013, brodu-arxiv-2014}. Polyacrylamide hydrogel is also sensitive to both temperature and dissolved salt concentration. This allows for precise system-wide tuning of the size and stiffness of the particles; it also enables studies of the dynamics of swelling or shrinking in response to local or global stimuli, which has particular relevance to biophysical systems~\cite{charras-nature-2005, lai-jbiomecheng-1991, dumais-annrevfluidmech-2012}.

In much the same way that granular monolayers and rafts of bubbles have served as indispensable model systems for developing fundamental concepts in the mechanics of both crystalline and amorphous materials~\cite{bragg-procrsoca-1947, argon-matscieng-1979, skjeltorp-nature-1988, meakin-science-1991}, so too can packings of soft particles provide unique insight into the deformation and failure of materials under nontrivial poromechanical loading, from the propagation of poroelastic waves to the coupling of deformation with flow and transport. In addition to serving as a tool for benchmarking numerical simulations~\cite{holtzman-pre-2010, zhang-arma-2011, zhang-jgr-2013}, this system offers an avenue into the experimental exploration of other fundamental problems of poroelasticity that have previously existed only as theoretical predictions or inferences from field observations.

\begin{acknowledgments}
    The authors gratefully acknowledge support from the Yale Climate \& Energy Institute. CWM also thanks T.~Bertrand, R.~F.~Katz, R.~W. Style, and L.~A.~Wilen for helpful discussions. ERD acknowledges support from the National Science Foundation (CBET-1236086). JSW acknowledges support from Yale~University, the Swedish Research Council (Vetenskapsr{\r{a}}det), and a Royal Society Wolfson Research Merit Award.
\end{acknowledgments}

\appendix
\section{Materials \& Methods}

\paragraph{Apparatus.} We pack a single layer of about 25,000 spherical particles (polyacrylamide hydrogel, JRM Chemical) between two borosilicate glass discs. The particles have mean diameter $d$$=$1.19~mm with standard deviation (SD) 0.12~mm (10\% polydispersity; Fig.~5). The discs are 19~mm thick and 212.7~mm in diameter, and they are separated by a plastic spacer that defines a working area of diameter $b$$=$210.5~mm and thickness $h$$=$1.44~mm. This is two SDs greater than $d$, confining the particles to a two-dimensional monolayer without restricting their in-plane motion. The spacer is permeable, allowing fluid to exit while confining the particles. We inject fluid at a fixed volume flow rate using a syringe pump (New Era NE-4000). The fluid is a mixture of water and glycerol (61\% glycerol by mass) with a viscosity of $\mu$$=$0.012~Pa$\cdot$s at 20~$^\circ{}$C. We acquire images with a digital camera (Canon EOS Rebel T2i).

\begin{figure}
   \centering
   \includegraphics[width=8.6cm]{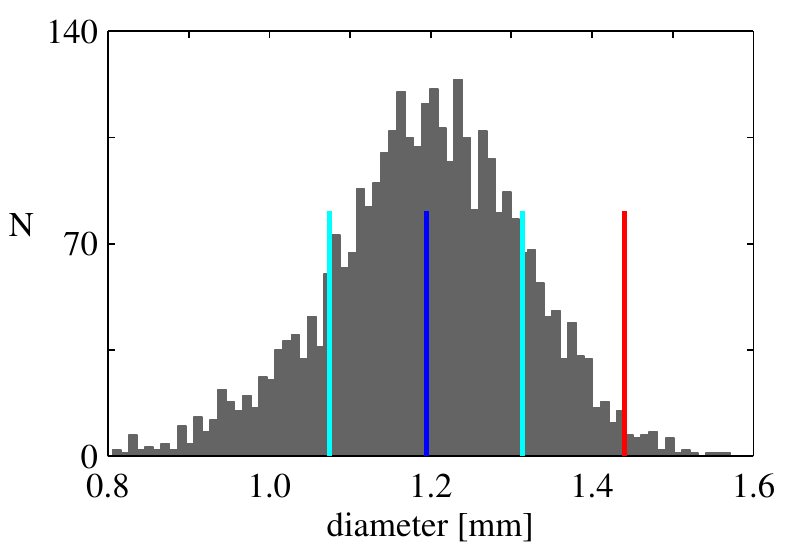}
   \caption{The particles have mean diameter 1.19~mm with standard deviation (SD) 0.12~mm (10\% polydispersity). Here, we show the particle-size distribution with the mean (vertical blue line) and the mean plus and minus one SD (vertical cyan lines). The gap between the glass discs is 1.44~mm (vertical red line), or two SDs above the mean, confining the particles to a monolayer without restricting their in-plane motion. This histogram was measured optically from a sample of about 3,230 particles. \label{fig:granulometry} }
\end{figure}

\paragraph{Particle detection and tracking.} We process the experimental images in \verb+MATLAB+, detecting particle positions in each image via centroid-finding after applying an intensity threshold. We track the particles from image to image using a standard particle-tracking algorithm~\cite{blair-code}. The images have a resolution of about 7.9~pixels per 1~mm (about 9.4~pixels per particle diameter). Frame-to-frame detection noise for particle centers is about 0.1 pixels (13~$\mu$m or about 1\% of one diameter; Figure~6).

\begin{figure}
   \centering
   \includegraphics[width=8.6cm]{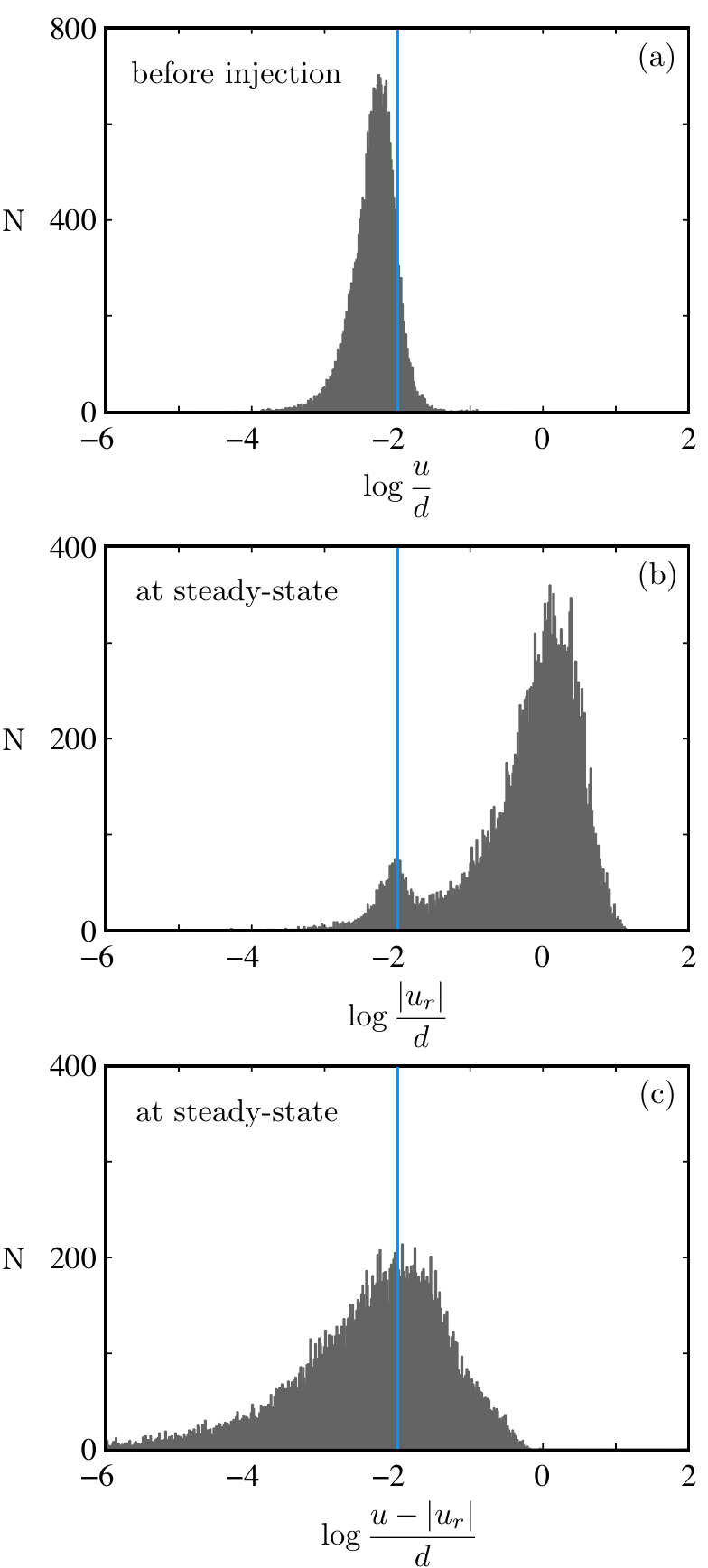}
   \caption{We detect particle positions to within about $\sim$0.01$d$. We estimate this detection noise by calculating (a)~total particle displacement between two frames taken before injection has started, when the particles are at rest. The noise has a mean of about 0.006$d$ particle diameters with a standard deviation of about 0.005$d$, indicating a detection threshold of $\sim$0.01$d$ (vertical blue line). At steady-state deformation, (b)~the average radial displacement is about $\sim$1.4$d$ with a standard deviation of about 1.6$d$, whereas (c)~the average non-radial displacement is about 0.03$d$ with a standard deviation of 0.06$d$, indicating that particle motion is almost entirely radial. \label{fig:ur_hist} }
\end{figure}

\paragraph{Deformation field.} We use the particle positions to calculate a best-fit local strain field following \cite{falk-pre-1998}. However, the quantity described in \cite{falk-pre-1998} as a local strain tensor $\boldsymbol{\epsilon}$ is more correctly identified as a local displacement gradient tensor $\partial{\boldsymbol{u}}/\partial{\boldsymbol{X}}$, where $\boldsymbol{X}$ is the undeformed configuration and $\boldsymbol{u}$ is the displacement field. The distinction is important here, where the displacements are large. We calculate the displacement gradient tensor and then use it to calculate the Green-Lagrange strain tensor $\boldsymbol{E}=\frac{1}{2}(\boldsymbol{F}^\intercal{}\boldsymbol{F}-\boldsymbol{I})$, where $\boldsymbol{F}=\boldsymbol{I}+\partial{\boldsymbol{u}}/\partial{\boldsymbol{X}}$ is the deformation gradient tensor and $\boldsymbol{I}$ is the identity tensor. The strains we report above are Green-Lagrange strains. We also calculate the root-mean-square (RMS) non-affine displacement $\bar{\boldsymbol{D}}_\mathrm{min}$ from their quantity $\boldsymbol{D}^2_\mathrm{min}$ by dividing each local value by the number of neighbors to form a mean (their quantity is a sum) and then taking the square root. The result has dimensions of length.

\section{Continuum model}

\subsection{Derivation}

We derive a continuum model for this process based on the theory of poroelasticity~\cite{coussy-wiley-2004}. We assume that the fluid and the solid are incompressible which, for the solid, implies that the beads can rearrange and deform without changing volume. This is a common assumption, justified here by the low working pressure ($\sim$5~kPa). Macroscopic deformation occurs through rearrangement of the solid skeleton, which leads to variations in the porosity (void or fluid fraction), $\phi_f$. Assuming axisymmetry and working strictly in terms of Eulerian quantities, conservation of mass dictates that
\begin{subequations}
    \begin{align}
        \frac{\partial{\phi_f}}{\partial{t}}+ \frac{1}{r}\frac{\partial}{\partial{r}}\big(r\phi_f{}v_f\big)&=0, \label{eq:continuity_f} \\
        \phi_f{}v_f+(1-\phi_f{})v_s&=\frac{Q}{2\pi{}rh}, \label{eq:macro-mass}
    \end{align}
\end{subequations}
where $v_f(r,t)$ and $v_s(r,t)$ are the velocity of the fluid and of the solid, $Q$ is the volume rate of fluid injection, and $h$ is the thickness of the gap. We assume that fluid flows relative to the solid skeleton according to Darcy's law,
\begin{equation}
	\phi_f{}(v_f-v_s)=-\frac{k}{\mu}\frac{\partial{p}}{\partial{r}}, \label{eq:darcy}
\end{equation}
where $k$ and $\mu$ are the permeability of the solid skeleton and the viscosity of the fluid, respectively, and $p(r,t)$ is the fluid pressure. Poroelastic theory dictates that the internal gradient in fluid pressure acts as a body force on the solid skeleton, and mechanical equilibrium requires that this must be supported by the divergence of the stress in the solid skeleton. We expect that the packing has an extremely low ratio of shear to bulk modulus, so for simplicity we assume that the solid cannot support shear or tensile stresses. This implies that the stress tensor is isotropic, and we write mechanical equilibrium as
\begin{equation}
	\nabla\cdot\sigma^\prime=\nabla{}p\quad\to\quad\frac{\partial{\sigma^\prime}}{\partial{r}}= \frac{\partial{p}}{\partial{r}}, \label{eq:equilibrium} 
\end{equation}
where $\sigma^\prime$ is the effective stress (\textit{i.e.}, the stress in the solid skeleton). Combining all of the above, we obtain a conservation law for the evolution of the porosity,
\begin{equation}\label{eq:model}
    \frac{\partial{\tilde{\phi}_f}}{\partial{t}}+\frac{1}{r}\frac{\partial}{\partial{r}} \bigg(\frac{Q}{2\pi{}h}\tilde{\phi}_f-r(1-\tilde{\phi}_f) \frac{k}{\mu}\frac{\partial{\sigma^\prime}}{\partial{r}}\bigg)=0,
\end{equation}
where $\tilde{\phi}_f=(\phi_f-\phi_{f,0})/(1-\phi_{f,0})$ is the Eulerian volumetric strain, or the normalized change in porosity, and $\phi_{f,0}$ is the initial (relaxed) porosity.

We must now specify a constitutive relationship between effective stress and strain, but note that Equation~\eqref{eq:model} is valid for any stress-strain relationship that yields an isotropic stress tensor. Here, we take the stress to be Hertzian elastic and linearly viscous in the volumetric strain. This can be written as
\begin{equation}\label{eq:sigmaprime}
	\sigma^\prime(\phi_f)=\mathcal{K}\tilde{\phi}_f|\tilde{\phi_f}|^{1/2}+\eta\frac{\partial{\tilde{\phi}_f}}{\partial{t}},
\end{equation}
where $\mathcal{K}$ and $\eta$ are the effective bulk modulus and viscosity of the solid skeleton, respectively. Note that the stress vanishes at $\phi_f=\phi_{f,0}\,\to\,\tilde{\phi}_f=0$. The linear viscous term introduces a simple ``rearrangement'' timescale, without which the mechanics would be quasi-static.

We assume that fluid is injected into a cavity in the solid of radius $a(t)$ and initial radius $a(0)=a_0$. The behavior is independent of $a_0$ for $a_0\ll{}b$. The fluid and the solid are initially at rest, $v_f(r,0)=v_s(r,0)=0$, and the initial porosity field is $\phi_f(r,0)=\phi_{f,0}\,\to\,\tilde{\phi}_f(r,0)=0$. Injection begins at $t=0$.

At the inner boundary, $r=a$, the normal component of the effective stress must vanish since the cavity is a free surface of the solid skeleton. For an isotropic stress field, this implies that $\sigma^\prime(a,t)=0$ and therefore that
\begin{equation}\label{eq:BCL}
    \tilde{\phi}_f(a,t)=0
\end{equation}
for $\tilde{\phi}_f(r,0)=0$. At the rigid outer boundary, $r=b$, we have that $u_r(b,t)=v_s(b,t)=0$. This can be written
\begin{equation}\label{eq:BCR}
    \frac{k}{\mu}\frac{\partial{\sigma^\prime}}{\partial{r}}\bigg|_{r=b} =-\frac{Q}{2\pi{}bh}.
\end{equation}
We also require an evolution equation for the position of the moving inner boundary,
\begin{equation}\label{eq:v_s_at_a}
    \frac{\mathrm{d}a}{\mathrm{d}t}=v_s(a,t)=\frac{Q}{2\pi{}ah}+ \frac{k}{\mu}\frac{\partial{\sigma^\prime}}{\partial{r}}\bigg|_{r=a}.
\end{equation}
Equations~\eqref{eq:model}--\eqref{eq:v_s_at_a} constitute a one-dimensional, time-dependent, moving-boundary problem. The solid displacement field does not appear explicitly, but we can calculate it at any time through a kinematic relationship,
\begin{equation}\label{eq:J_to_phi}
    J=\det{\boldsymbol{F}}=\frac{1-\phi_{f,0}}{1-\phi_f},
\end{equation}
where $J$ is the Jacobian determinant, or the Jacobian of the deformation, and $\boldsymbol{F}$ is the deformation gradient tensor~(see Appendix~A). For an axisymmetric, plane-strain deformation, Eq.~\eqref{eq:J_to_phi} becomes
\begin{equation}\label{eq:phi_to_u}
    \frac{\phi_f-\phi_{f,0}}{1-\phi_{f,0}}=\tilde{\phi}_f=\frac{1}{r}\frac{\partial}{\partial{r}} \left(ru_r-\frac{1}{2}u_r^2\right),
\end{equation}
where $u_r(r,t)$ is the $r$-component of the (Eulerian) solid displacement field. In contrast with linear poroelasticity, this problem is nonlinear for four reasons: We account rigorously for (1)~the moving boundary, (2)~the solid velocity, and (3)~the exact relationship between porosity and displacement, and we use (4)~a nonlinear elastic law.

A similar axisymmetric model was developed by \cite{barry-jaustralmathsocb-1993} for small elastic deformations. Our model incorporates a nonlinear elastic law and linear viscous effects, and is kinematically exact---it is valid for arbitrarily large deformations as long as the constitutive laws remain valid.

\subsection{Dimensionless form}

We present Equations~\eqref{eq:model} and \eqref{eq:sigmaprime} in dimensionless form in the main text (Equations~\ref{eq:model_nd} and \ref{eq:sigmaprime_nd}, respectively), where $\tilde{r}=r/b$ is the radial coordinate scaled by the outer radius, $\tilde{t}=t/T_\mathrm{pe}$ is time scaled by the poroelastic time scale, $\tilde{\sigma}^\prime=\sigma^\prime/\mathcal{K}$ is the effective stress scaled by the bulk modulus, and the two dimensionless parameters are $\alpha=\mu{}Q/(2\pi{}h\mathcal{K}k)$ and $\beta=\eta{}k/(\mu{}b^2)$. The dimensionless boundary conditions are
\begin{equation}
    \tilde{\phi}_f(\tilde{a},\tilde{t})=0,
\end{equation}
\begin{equation}
    \frac{\partial{\tilde{\sigma}^\prime}}{\partial{\tilde{r}}}\bigg|_{\tilde{r}=1}=-\alpha,
\end{equation}
and
\begin{equation}
    \frac{\mathrm{d}\tilde{a}}{\mathrm{d}\tilde{t}}=\frac{\alpha}{\tilde{a}}+ \frac{\partial{\tilde{\sigma}^\prime}}{\partial{\tilde{r}}}\bigg|_{\tilde{r}=\tilde{a}},
\end{equation}
where $\tilde{a}=a/b$. We solve the model numerically using a finite-volume method with explicit time integration, accommodating the moving boundary with an adaptive grid. At steady-state, the model has implicit solution
\begin{equation}
    \tilde{\phi}_{f,ss}=-\left[\alpha\ln(\tilde{r}/\tilde{a}_{ss})\right]^{2/3}
\end{equation}
and
\begin{equation}
    \tilde{u}_{r,ss}(\tilde{r})=\tilde{r}-\sqrt{\tilde{r}^2-2\mathcal{I}},
\end{equation}
where the integral $\mathcal{I}(\tilde{r})$ is
\begin{equation}
        \mathcal{I}(\tilde{r}) =-\int_{\tilde{r}}^1\,r\tilde{\phi}_{f,ss}(r)\,\mathrm{d}r =\int_{\tilde{r}}^1\, r\left[\alpha{}\ln(r/\tilde{a}_{ss})\right]^{2/3}\,\mathrm{d}r \label{eq:I_of_r}
\end{equation}
and the subscript ``$ss$'' refers to the value of a quantity at steady state. The steady-state cavity radius is determined implicitly by the definition $\tilde{u}_{r,ss}(\tilde{a}_{ss})=\tilde{a}_{ss}-\tilde{a}_0$, which leads to
\begin{equation}\label{eq:aroot}
    \tilde{a}_{ss}=\sqrt{\tilde{a}_0^2+2\mathcal{I}(\tilde{a}_{ss})}.
\end{equation}
We solve Equation~\eqref{eq:aroot} numerically using a standard root-finding technique, evaluating the integral $\mathcal{I}(\tilde{a}_{ss})$ from Equation~\eqref{eq:I_of_r} using the trapezoidal rule.

The mechanical properties of the packing are difficult to measure and strongly dependent on the particle arrangement~(Figure~2c), so we use the mechanical properties as fitting parameters. To compare the model with the experiment at steady-state, as in Figure~4, we choose a value of the product $\mathcal{K}k$ to match the steady-state cavity radius. We use a single value of $\mathcal{K}k$ for all of the experiments in Figure~2c. To compare the dynamics, as in Figures~2a and 7, we choose $\mathcal{K}k$ to match the steady-state cavity radius and then $\eta{}k$ to match the rate of deformation and relaxation ($\eta{}k$ plays no role in the steady state).

\begin{figure}
   \centering
   \includegraphics[width=8.6cm]{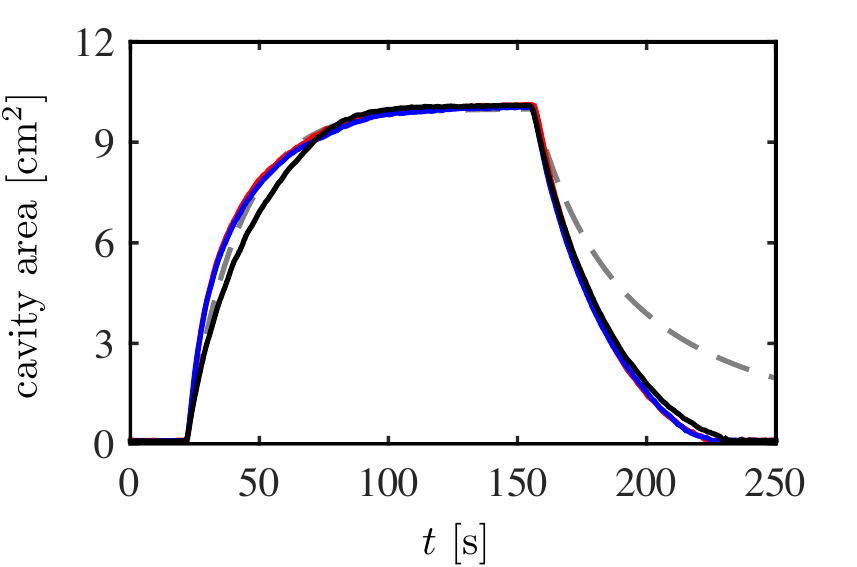}
   \caption{As Figure~2a of the main text, but without the 12~s offset. The second and third cycles (blue and red) and the model (dashed gray) agree very well during cavity expansion, but the model relaxes more slowly than the experiment. The first cycle of the experiment (black) is somewhat different due to heterogeneities in the initial condition. \label{fig:cavity_supp} }
\end{figure}


\begin{thebibliography}{57}%
\makeatletter
\providecommand \@ifxundefined [1]{%
 \@ifx{#1\undefined}
}%
\providecommand \@ifnum [1]{%
 \ifnum #1\expandafter \@firstoftwo
 \else \expandafter \@secondoftwo
 \fi
}%
\providecommand \@ifx [1]{%
 \ifx #1\expandafter \@firstoftwo
 \else \expandafter \@secondoftwo
 \fi
}%
\providecommand \natexlab [1]{#1}%
\providecommand \enquote  [1]{``#1''}%
\providecommand \bibnamefont  [1]{#1}%
\providecommand \bibfnamefont [1]{#1}%
\providecommand \citenamefont [1]{#1}%
\providecommand \href@noop [0]{\@secondoftwo}%
\providecommand \href [0]{\begingroup \@sanitize@url \@href}%
\providecommand \@href[1]{\@@startlink{#1}\@@href}%
\providecommand \@@href[1]{\endgroup#1\@@endlink}%
\providecommand \@sanitize@url [0]{\catcode `\\12\catcode `\$12\catcode
  `\&12\catcode `\#12\catcode `\^12\catcode `\_12\catcode `\%12\relax}%
\providecommand \@@startlink[1]{}%
\providecommand \@@endlink[0]{}%
\providecommand \url  [0]{\begingroup\@sanitize@url \@url }%
\providecommand \@url [1]{\endgroup\@href {#1}{\urlprefix }}%
\providecommand \urlprefix  [0]{URL }%
\providecommand \Eprint [0]{\href }%
\providecommand \doibase [0]{http://dx.doi.org/}%
\providecommand \selectlanguage [0]{\@gobble}%
\providecommand \bibinfo  [0]{\@secondoftwo}%
\providecommand \bibfield  [0]{\@secondoftwo}%
\providecommand \translation [1]{[#1]}%
\providecommand \BibitemOpen [0]{}%
\providecommand \bibitemStop [0]{}%
\providecommand \bibitemNoStop [0]{.\EOS\space}%
\providecommand \EOS [0]{\spacefactor3000\relax}%
\providecommand \BibitemShut  [1]{\csname bibitem#1\endcsname}%
\let\auto@bib@innerbib\@empty
\bibitem [{\citenamefont {{von Terzaghi}}(1936)}]{terzaghi-procsmfe-1936}%
  \BibitemOpen
  \bibfield  {author} {\bibinfo {author} {\bibfnamefont {K.}~\bibnamefont {{von
  Terzaghi}}},\ }\bibfield  {title} {\enquote {\bibinfo {title} {The shearing
  resistance of saturated soil and the angle between the planes of shear},}\
  }in\ \href@noop {} {\emph {\bibinfo {booktitle} {Proceedings of the
  International Conference on Soil Mechanics and Foundation Engineering, June
  22 to 26}}},\ Vol.~\bibinfo {volume} {1}\ (\bibinfo {year} {1936})\ pp.\
  \bibinfo {pages} {54--56}\BibitemShut {NoStop}%
\bibitem [{\citenamefont {Biot}(1941)}]{biot-jap-1941}%
  \BibitemOpen
  \bibfield  {author} {\bibinfo {author} {\bibfnamefont {M.~A.}\ \bibnamefont
  {Biot}},\ }\bibfield  {title} {\enquote {\bibinfo {title} {General theory of
  three-dimensional consolidation},}\ }\href@noop {} {\bibfield  {journal}
  {\bibinfo  {journal} {Journal of Applied Physics}\ }\textbf {\bibinfo
  {volume} {12}},\ \bibinfo {pages} {155--164} (\bibinfo {year}
  {1941})}\BibitemShut {NoStop}%
\bibitem [{\citenamefont {Atkin}\ and\ \citenamefont
  {Craine}(1976)}]{atkin-imajapplmath-1976}%
  \BibitemOpen
  \bibfield  {author} {\bibinfo {author} {\bibfnamefont {R.~J.}\ \bibnamefont
  {Atkin}}\ and\ \bibinfo {author} {\bibfnamefont {R.~E.}\ \bibnamefont
  {Craine}},\ }\bibfield  {title} {\enquote {\bibinfo {title} {Continuum
  theories of mixtures: {Applications}},}\ }\href {\doibase
  10.1093/imamat/17.2.153} {\bibfield  {journal} {\bibinfo  {journal} {{IMA}
  Journal of Applied Mathematics}\ }\textbf {\bibinfo {volume} {17}},\ \bibinfo
  {pages} {153--207} (\bibinfo {year} {1976})}\BibitemShut {NoStop}%
\bibitem [{\citenamefont {Kenyon}(1976)}]{kenyon-archrationmechanalysis-1976b}%
  \BibitemOpen
  \bibfield  {author} {\bibinfo {author} {\bibfnamefont {D.~E.}\ \bibnamefont
  {Kenyon}},\ }\bibfield  {title} {\enquote {\bibinfo {title} {The theory of an
  incompressible solid-fluid mixture},}\ }\href {\doibase 10.1007/BF00248468}
  {\bibfield  {journal} {\bibinfo  {journal} {Archive for Rational Mechanics
  and Analysis}\ }\textbf {\bibinfo {volume} {62}},\ \bibinfo {pages}
  {131--147} (\bibinfo {year} {1976})}\BibitemShut {NoStop}%
\bibitem [{\citenamefont {Coussy}(2004)}]{coussy-wiley-2004}%
  \BibitemOpen
  \bibfield  {author} {\bibinfo {author} {\bibfnamefont {O.}~\bibnamefont
  {Coussy}},\ }\href@noop {} {\emph {\bibinfo {title} {Poromechanics}}}\
  (\bibinfo  {publisher} {Wiley},\ \bibinfo {year} {2004})\BibitemShut
  {NoStop}%
\bibitem [{\citenamefont {Yang}\ and\ \citenamefont
  {Taber}(1991)}]{yang-jbiomech-1991}%
  \BibitemOpen
  \bibfield  {author} {\bibinfo {author} {\bibfnamefont {M.}~\bibnamefont
  {Yang}}\ and\ \bibinfo {author} {\bibfnamefont {L.~A.}\ \bibnamefont
  {Taber}},\ }\bibfield  {title} {\enquote {\bibinfo {title} {The possible role
  of poroelasticity in the apparent viscoelastic behavior of passive cardiac
  muscle},}\ }\href {\doibase 10.1016/0021-9290(91)90291-T} {\bibfield
  {journal} {\bibinfo  {journal} {Journal of Biomechanics}\ }\textbf {\bibinfo
  {volume} {24}},\ \bibinfo {pages} {587--597} (\bibinfo {year}
  {1991})}\BibitemShut {NoStop}%
\bibitem [{\citenamefont {Lai}\ \emph {et~al.}(1991)\citenamefont {Lai},
  \citenamefont {Hou},\ and\ \citenamefont {Mow}}]{lai-jbiomecheng-1991}%
  \BibitemOpen
  \bibfield  {author} {\bibinfo {author} {\bibfnamefont {W.~M.}\ \bibnamefont
  {Lai}}, \bibinfo {author} {\bibfnamefont {J.~S.}\ \bibnamefont {Hou}}, \ and\
  \bibinfo {author} {\bibfnamefont {V.~C.}\ \bibnamefont {Mow}},\ }\bibfield
  {title} {\enquote {\bibinfo {title} {A triphasic theory for the swelling and
  deformation behaviors of articular cartilage},}\ }\href {\doibase
  10.1115/1.2894880} {\bibfield  {journal} {\bibinfo  {journal} {Journal of
  Biomechanical Engineering}\ }\textbf {\bibinfo {volume} {113}},\ \bibinfo
  {pages} {245--258} (\bibinfo {year} {1991})}\BibitemShut {NoStop}%
\bibitem [{\citenamefont {Cowin}(1999)}]{cowin-jbiomech-1999}%
  \BibitemOpen
  \bibfield  {author} {\bibinfo {author} {\bibfnamefont {S.~C.}\ \bibnamefont
  {Cowin}},\ }\bibfield  {title} {\enquote {\bibinfo {title} {Bone
  poroelasticity},}\ }\href {\doibase 10.1016/S0021-9290(98)00161-4} {\bibfield
   {journal} {\bibinfo  {journal} {Journal of Biomechanics}\ }\textbf {\bibinfo
  {volume} {32}},\ \bibinfo {pages} {217--238} (\bibinfo {year}
  {1999})}\BibitemShut {NoStop}%
\bibitem [{\citenamefont {Charras}\ \emph {et~al.}(2005)\citenamefont
  {Charras}, \citenamefont {Yarrow}, \citenamefont {Horton}, \citenamefont
  {Mahadevan},\ and\ \citenamefont {Mitchison}}]{charras-nature-2005}%
  \BibitemOpen
  \bibfield  {author} {\bibinfo {author} {\bibfnamefont {G.~T.}\ \bibnamefont
  {Charras}}, \bibinfo {author} {\bibfnamefont {J.~C.}\ \bibnamefont {Yarrow}},
  \bibinfo {author} {\bibfnamefont {M.~A.}\ \bibnamefont {Horton}}, \bibinfo
  {author} {\bibfnamefont {L.}~\bibnamefont {Mahadevan}}, \ and\ \bibinfo
  {author} {\bibfnamefont {T.~J.}\ \bibnamefont {Mitchison}},\ }\bibfield
  {title} {\enquote {\bibinfo {title} {Non-equilibration of hydrostatic
  pressure in blebbing cells},}\ }\href {\doibase 10.1038/nature03550}
  {\bibfield  {journal} {\bibinfo  {journal} {Nature}\ }\textbf {\bibinfo
  {volume} {435}},\ \bibinfo {pages} {365--369} (\bibinfo {year}
  {2005})}\BibitemShut {NoStop}%
\bibitem [{\citenamefont {Moeendarbary}\ \emph {et~al.}(2013)\citenamefont
  {Moeendarbary}, \citenamefont {Valon}, \citenamefont {Fritzsche},
  \citenamefont {Harris}, \citenamefont {Moulding}, \citenamefont {Thrasher},
  \citenamefont {Stride}, \citenamefont {Mahadevan},\ and\ \citenamefont
  {Charras}}]{moeendarbary-natmaterials-2013}%
  \BibitemOpen
  \bibfield  {author} {\bibinfo {author} {\bibfnamefont {E.}~\bibnamefont
  {Moeendarbary}}, \bibinfo {author} {\bibfnamefont {L.}~\bibnamefont {Valon}},
  \bibinfo {author} {\bibfnamefont {M.}~\bibnamefont {Fritzsche}}, \bibinfo
  {author} {\bibfnamefont {A.~R.}\ \bibnamefont {Harris}}, \bibinfo {author}
  {\bibfnamefont {D.~A.}\ \bibnamefont {Moulding}}, \bibinfo {author}
  {\bibfnamefont {A.~J.}\ \bibnamefont {Thrasher}}, \bibinfo {author}
  {\bibfnamefont {E.}~\bibnamefont {Stride}}, \bibinfo {author} {\bibfnamefont
  {L.}~\bibnamefont {Mahadevan}}, \ and\ \bibinfo {author} {\bibfnamefont
  {G.~T.}\ \bibnamefont {Charras}},\ }\bibfield  {title} {\enquote {\bibinfo
  {title} {The cytoplasm of living cells behaves as a poroelastic material},}\
  }\href {\doibase 10.1038/nmat3517} {\bibfield  {journal} {\bibinfo  {journal}
  {Nature Materials}\ }\textbf {\bibinfo {volume} {12}},\ \bibinfo {pages}
  {253--261} (\bibinfo {year} {2013})}\BibitemShut {NoStop}%
\bibitem [{\citenamefont {Dumais}\ and\ \citenamefont
  {Forterre}(2012)}]{dumais-annrevfluidmech-2012}%
  \BibitemOpen
  \bibfield  {author} {\bibinfo {author} {\bibfnamefont {J.}~\bibnamefont
  {Dumais}}\ and\ \bibinfo {author} {\bibfnamefont {Y.}~\bibnamefont
  {Forterre}},\ }\bibfield  {title} {\enquote {\bibinfo {title} {``{Vegetable
  Dynamicks}'': {T}he role of water in plant movements},}\ }\href {\doibase
  10.1146/annurev-fluid-120710-101200} {\bibfield  {journal} {\bibinfo
  {journal} {Annual Review of Fluid Mechanics}\ }\textbf {\bibinfo {volume}
  {44}},\ \bibinfo {pages} {453--478} (\bibinfo {year} {2012})}\BibitemShut
  {NoStop}%
\bibitem [{\citenamefont {Szulczewski}\ \emph {et~al.}(2012)\citenamefont
  {Szulczewski}, \citenamefont {MacMinn}, \citenamefont {Herzog},\ and\
  \citenamefont {Juanes}}]{szulczewski-pnas-2012}%
  \BibitemOpen
  \bibfield  {author} {\bibinfo {author} {\bibfnamefont {M.~L.}\ \bibnamefont
  {Szulczewski}}, \bibinfo {author} {\bibfnamefont {C.~W.}\ \bibnamefont
  {MacMinn}}, \bibinfo {author} {\bibfnamefont {H.~J.}\ \bibnamefont {Herzog}},
  \ and\ \bibinfo {author} {\bibfnamefont {R.}~\bibnamefont {Juanes}},\
  }\bibfield  {title} {\enquote {\bibinfo {title} {Lifetime of carbon capture
  and storage as a climate-change mitigation technology},}\ }\href {\doibase
  10.1073/pnas.1115347109} {\bibfield  {journal} {\bibinfo  {journal}
  {Proceedings of the National Academy of Sciences of the United States of
  America}\ }\textbf {\bibinfo {volume} {109}},\ \bibinfo {pages} {5185--5189}
  (\bibinfo {year} {2012})}\BibitemShut {NoStop}%
\bibitem [{\citenamefont {{National Research
  Council}}(2013)}]{nrc-nationalacademiespress-2013}%
  \BibitemOpen
  \bibfield  {author} {\bibinfo {author} {\bibnamefont {{National Research
  Council}}},\ }\href@noop {} {\emph {\bibinfo {title} {Induced Seismicity
  Potential in Energy Technologies}}}\ (\bibinfo  {publisher} {The National
  Academies Press},\ \bibinfo {address} {Washington, DC},\ \bibinfo {year}
  {2013})\BibitemShut {NoStop}%
\bibitem [{\citenamefont {Verdon}\ \emph {et~al.}(2013)\citenamefont {Verdon},
  \citenamefont {Kendall}, \citenamefont {Stork}, \citenamefont {Chadwick},
  \citenamefont {White},\ and\ \citenamefont {Bissell}}]{verdon-pnas-2013}%
  \BibitemOpen
  \bibfield  {author} {\bibinfo {author} {\bibfnamefont {J.~P.}\ \bibnamefont
  {Verdon}}, \bibinfo {author} {\bibfnamefont {J.-M.}\ \bibnamefont {Kendall}},
  \bibinfo {author} {\bibfnamefont {A.~L.}\ \bibnamefont {Stork}}, \bibinfo
  {author} {\bibfnamefont {R.~A.}\ \bibnamefont {Chadwick}}, \bibinfo {author}
  {\bibfnamefont {D.~J.}\ \bibnamefont {White}}, \ and\ \bibinfo {author}
  {\bibfnamefont {R.~C.}\ \bibnamefont {Bissell}},\ }\bibfield  {title}
  {\enquote {\bibinfo {title} {Comparison of geomechanical deformation induced
  by megatonne-scale {CO}$_2$ storage at {Sleipner}, {Weyburn}, and {In
  Salah}},}\ }\href {\doibase 10.1073/pnas.1302156110} {\bibfield  {journal}
  {\bibinfo  {journal} {Proceedings of the National Academy of Sciences of the
  United States of America}\ }\textbf {\bibinfo {volume} {110}},\ \bibinfo
  {pages} {E2762--71} (\bibinfo {year} {2013})}\BibitemShut {NoStop}%
\bibitem [{\citenamefont {Jha}\ and\ \citenamefont
  {Juanes}(2014)}]{jha-wrr-2014}%
  \BibitemOpen
  \bibfield  {author} {\bibinfo {author} {\bibfnamefont {B.}~\bibnamefont
  {Jha}}\ and\ \bibinfo {author} {\bibfnamefont {R.}~\bibnamefont {Juanes}},\
  }\bibfield  {title} {\enquote {\bibinfo {title} {Coupled multiphase flow and
  poromechanics: {A} computational model of pore-pressure effects on fault slip
  and earthquake triggering},}\ }\href {\doibase 10.1002/2013WR015175}
  {\bibfield  {journal} {\bibinfo  {journal} {Water Resources Research}\
  }\textbf {\bibinfo {volume} {50}},\ \bibinfo {pages} {3776--3808} (\bibinfo
  {year} {2014})}\BibitemShut {NoStop}%
\bibitem [{\citenamefont {Hubbert}\ and\ \citenamefont
  {Willis}(1957)}]{hubbert-transaime-1957}%
  \BibitemOpen
  \bibfield  {author} {\bibinfo {author} {\bibfnamefont {M.~K.}\ \bibnamefont
  {Hubbert}}\ and\ \bibinfo {author} {\bibfnamefont {D.~G.}\ \bibnamefont
  {Willis}},\ }\bibfield  {title} {\enquote {\bibinfo {title} {Mechanics of
  hydraulic fracturing},}\ }\href@noop {} {\bibfield  {journal} {\bibinfo
  {journal} {Petroleum Transactions, American Institute of Mining,
  Metallurgical, and Petroleum Engineers}\ }\textbf {\bibinfo {volume} {210}},\
  \bibinfo {pages} {153--168} (\bibinfo {year} {1957})},\ \bibinfo {note}
  {number SPE-686-G}\BibitemShut {NoStop}%
\bibitem [{\citenamefont {Detournay}\ and\ \citenamefont
  {Cheng}(1988)}]{detournay-intjrmms-1988}%
  \BibitemOpen
  \bibfield  {author} {\bibinfo {author} {\bibfnamefont {E.}~\bibnamefont
  {Detournay}}\ and\ \bibinfo {author} {\bibfnamefont {A.~H.-D.}\ \bibnamefont
  {Cheng}},\ }\bibfield  {title} {\enquote {\bibinfo {title} {Poroelastic
  response of a borehole in a non-hydrostatic stress field},}\ }\href {\doibase
  10.1016/0148-9062(88)92299-1} {\bibfield  {journal} {\bibinfo  {journal}
  {International Journal of Rock Mechanics and Mining Sciences \&
  Geomechanics}\ }\textbf {\bibinfo {volume} {25}},\ \bibinfo {pages}
  {171--182} (\bibinfo {year} {1988})}\BibitemShut {NoStop}%
\bibitem [{\citenamefont {Yarushina}\ \emph {et~al.}(2013)\citenamefont
  {Yarushina}, \citenamefont {Bercovici},\ and\ \citenamefont
  {Oristaglio}}]{yarushina-geophysjint-2013}%
  \BibitemOpen
  \bibfield  {author} {\bibinfo {author} {\bibfnamefont {V.~M.}\ \bibnamefont
  {Yarushina}}, \bibinfo {author} {\bibfnamefont {D.}~\bibnamefont
  {Bercovici}}, \ and\ \bibinfo {author} {\bibfnamefont {M.~L.}\ \bibnamefont
  {Oristaglio}},\ }\bibfield  {title} {\enquote {\bibinfo {title} {Rock
  deformation models and fluid leak-off in hydraulic fracturing},}\ }\href
  {\doibase 10.1093/gji/ggt199} {\bibfield  {journal} {\bibinfo  {journal}
  {Geophysical Journal International}\ }\textbf {\bibinfo {volume} {194}},\
  \bibinfo {pages} {1514--1526} (\bibinfo {year} {2013})}\BibitemShut {NoStop}%
\bibitem [{\citenamefont {Wang}(2000)}]{wang-princeton-2000}%
  \BibitemOpen
  \bibfield  {author} {\bibinfo {author} {\bibfnamefont {H.~F.}\ \bibnamefont
  {Wang}},\ }\href@noop {} {\emph {\bibinfo {title} {Theory of Linear
  Poroelasticity}}}\ (\bibinfo  {publisher} {Princeton University Press},\
  \bibinfo {address} {Princeton NJ},\ \bibinfo {year} {2000})\BibitemShut
  {NoStop}%
\bibitem [{\citenamefont {Yu}\ and\ \citenamefont
  {Houlsby}(1991)}]{yu-geotechnique-1991}%
  \BibitemOpen
  \bibfield  {author} {\bibinfo {author} {\bibfnamefont {H.~S.}\ \bibnamefont
  {Yu}}\ and\ \bibinfo {author} {\bibfnamefont {G.~T.}\ \bibnamefont
  {Houlsby}},\ }\bibfield  {title} {\enquote {\bibinfo {title} {Finite cavity
  expansion in dilatant soils: {Loading} analysis},}\ }\href {\doibase
  10.1680/geot.1991.41.2.173} {\bibfield  {journal} {\bibinfo  {journal}
  {G{\'{e}}otechnique}\ }\textbf {\bibinfo {volume} {41}},\ \bibinfo {pages}
  {173--183} (\bibinfo {year} {1991})}\BibitemShut {NoStop}%
\bibitem [{\citenamefont {Alsiny}\ \emph {et~al.}(1992)\citenamefont {Alsiny},
  \citenamefont {Vardoulakis},\ and\ \citenamefont
  {Drescher}}]{alsiny-geotechnique-1992}%
  \BibitemOpen
  \bibfield  {author} {\bibinfo {author} {\bibfnamefont {A.}~\bibnamefont
  {Alsiny}}, \bibinfo {author} {\bibfnamefont {I.}~\bibnamefont {Vardoulakis}},
  \ and\ \bibinfo {author} {\bibfnamefont {A.}~\bibnamefont {Drescher}},\
  }\bibfield  {title} {\enquote {\bibinfo {title} {Deformation localization in
  cavity inflation experiments on dry sand},}\ }\href {\doibase
  10.1680/geot.1992.42.3.395} {\bibfield  {journal} {\bibinfo  {journal}
  {G{\'{e}}otechnique}\ }\textbf {\bibinfo {volume} {42}},\ \bibinfo {pages}
  {395--410} (\bibinfo {year} {1992})}\BibitemShut {NoStop}%
\bibitem [{\citenamefont {Hutchens}\ and\ \citenamefont
  {Crosby}(2014)}]{hutchens-softmatter-2014}%
  \BibitemOpen
  \bibfield  {author} {\bibinfo {author} {\bibfnamefont {S.~B.}\ \bibnamefont
  {Hutchens}}\ and\ \bibinfo {author} {\bibfnamefont {A.~J.}\ \bibnamefont
  {Crosby}},\ }\bibfield  {title} {\enquote {\bibinfo {title} {Soft-solid
  deformation mechanics at the tip of an embedded needle},}\ }\href {\doibase
  10.1039/c3sm52689e} {\bibfield  {journal} {\bibinfo  {journal} {Soft Matter}\
  }\textbf {\bibinfo {volume} {10}},\ \bibinfo {pages} {3679--3684} (\bibinfo
  {year} {2014})}\BibitemShut {NoStop}%
\bibitem [{\citenamefont {Coulais}\ \emph {et~al.}(2014)\citenamefont
  {Coulais}, \citenamefont {Seguin},\ and\ \citenamefont
  {Dauchot}}]{coulais-prl-2014}%
  \BibitemOpen
  \bibfield  {author} {\bibinfo {author} {\bibfnamefont {C.}~\bibnamefont
  {Coulais}}, \bibinfo {author} {\bibfnamefont {A.}~\bibnamefont {Seguin}}, \
  and\ \bibinfo {author} {\bibfnamefont {O.}~\bibnamefont {Dauchot}},\
  }\bibfield  {title} {\enquote {\bibinfo {title} {Shear modulus and dilatancy
  softening in granular packings above jamming},}\ }\href {\doibase
  10.1103/PhysRevLett.113.198001} {\bibfield  {journal} {\bibinfo  {journal}
  {Physical Review Letters}\ }\textbf {\bibinfo {volume} {113}},\ \bibinfo
  {pages} {198001} (\bibinfo {year} {2014})}\BibitemShut {NoStop}%
\bibitem [{\citenamefont {Johnsen}\ \emph {et~al.}(2006)\citenamefont
  {Johnsen}, \citenamefont {Toussaint}, \citenamefont {M{\r{a}}l{\o}y},\ and\
  \citenamefont {Flekk{\o}y}}]{johnsen-pre-2006}%
  \BibitemOpen
  \bibfield  {author} {\bibinfo {author} {\bibfnamefont {{\O}.}~\bibnamefont
  {Johnsen}}, \bibinfo {author} {\bibfnamefont {R.}~\bibnamefont {Toussaint}},
  \bibinfo {author} {\bibfnamefont {K.~J.}\ \bibnamefont {M{\r{a}}l{\o}y}}, \
  and\ \bibinfo {author} {\bibfnamefont {E.~G.}\ \bibnamefont {Flekk{\o}y}},\
  }\bibfield  {title} {\enquote {\bibinfo {title} {Pattern formation during air
  injection into granular materials confined in a circular {H}ele-{S}haw
  cell},}\ }\href {\doibase 10.1103/PhysRevE.74.011301} {\bibfield  {journal}
  {\bibinfo  {journal} {Physical Review E}\ }\textbf {\bibinfo {volume} {74}},\
  \bibinfo {pages} {011301} (\bibinfo {year} {2006})}\BibitemShut {NoStop}%
\bibitem [{\citenamefont {Cheng}\ \emph {et~al.}(2008)\citenamefont {Cheng},
  \citenamefont {Xu}, \citenamefont {Patterson}, \citenamefont {Jaeger},\ and\
  \citenamefont {Nagel}}]{cheng-natphys-2008}%
  \BibitemOpen
  \bibfield  {author} {\bibinfo {author} {\bibfnamefont {X.}~\bibnamefont
  {Cheng}}, \bibinfo {author} {\bibfnamefont {L.}~\bibnamefont {Xu}}, \bibinfo
  {author} {\bibfnamefont {A.}~\bibnamefont {Patterson}}, \bibinfo {author}
  {\bibfnamefont {H.~M.}\ \bibnamefont {Jaeger}}, \ and\ \bibinfo {author}
  {\bibfnamefont {S.~R.}\ \bibnamefont {Nagel}},\ }\bibfield  {title} {\enquote
  {\bibinfo {title} {Towards the zero-surface-tension limit in granular
  fingering instability},}\ }\href {\doibase 10.1038/nphys834} {\bibfield
  {journal} {\bibinfo  {journal} {Nature Physics}\ }\textbf {\bibinfo {volume}
  {4}},\ \bibinfo {pages} {234--237} (\bibinfo {year} {2008})}\BibitemShut
  {NoStop}%
\bibitem [{\citenamefont {Sandnes}\ \emph {et~al.}(2011)\citenamefont
  {Sandnes}, \citenamefont {Flekk{{\o}}y}, \citenamefont {Knudsen},
  \citenamefont {M{\r{a}}l{\o}y},\ and\ \citenamefont
  {See}}]{sandnes-natcomms-2011}%
  \BibitemOpen
  \bibfield  {author} {\bibinfo {author} {\bibfnamefont {B.}~\bibnamefont
  {Sandnes}}, \bibinfo {author} {\bibfnamefont {E.~G.}\ \bibnamefont
  {Flekk{{\o}}y}}, \bibinfo {author} {\bibfnamefont {H.~A.}\ \bibnamefont
  {Knudsen}}, \bibinfo {author} {\bibfnamefont {K.~J.}\ \bibnamefont
  {M{\r{a}}l{\o}y}}, \ and\ \bibinfo {author} {\bibfnamefont {H.}~\bibnamefont
  {See}},\ }\bibfield  {title} {\enquote {\bibinfo {title} {Patterns and flow
  in frictional fluid dynamics},}\ }\href {\doibase 10.1038/ncomms1289}
  {\bibfield  {journal} {\bibinfo  {journal} {Nature Communications}\ }\textbf
  {\bibinfo {volume} {2}},\ \bibinfo {pages} {288} (\bibinfo {year} {2011})}\BibitemShut {NoStop}%
\bibitem [{\citenamefont {Huang}\ \emph {et~al.}(2012)\citenamefont {Huang},
  \citenamefont {Zhang}, \citenamefont {Callahan},\ and\ \citenamefont
  {Ayoub}}]{huang-prl-2012}%
  \BibitemOpen
  \bibfield  {author} {\bibinfo {author} {\bibfnamefont {H.}~\bibnamefont
  {Huang}}, \bibinfo {author} {\bibfnamefont {F.}~\bibnamefont {Zhang}},
  \bibinfo {author} {\bibfnamefont {P.}~\bibnamefont {Callahan}}, \ and\
  \bibinfo {author} {\bibfnamefont {J.}~\bibnamefont {Ayoub}},\ }\bibfield
  {title} {\enquote {\bibinfo {title} {Granular fingering of fluid injection
  into dense granular media in a {H}ele-{S}haw cell},}\ }\href {\doibase
  10.1103/PhysRevLett.108.258001} {\bibfield  {journal} {\bibinfo  {journal}
  {Physical Review Letters}\ }\textbf {\bibinfo {volume} {108}},\ \bibinfo
  {pages} {258001} (\bibinfo {year} {2012})}\BibitemShut {NoStop}%
\bibitem [{\citenamefont {Holtzman}\ \emph {et~al.}(2012)\citenamefont
  {Holtzman}, \citenamefont {Szulczewski},\ and\ \citenamefont
  {Juanes}}]{holtzman-prl-2012}%
  \BibitemOpen
  \bibfield  {author} {\bibinfo {author} {\bibfnamefont {R.}~\bibnamefont
  {Holtzman}}, \bibinfo {author} {\bibfnamefont {M.~L.}\ \bibnamefont
  {Szulczewski}}, \ and\ \bibinfo {author} {\bibfnamefont {R.}~\bibnamefont
  {Juanes}},\ }\bibfield  {title} {\enquote {\bibinfo {title} {Capillary
  fracturing in granular media},}\ }\href {\doibase
  10.1103/PhysRevLett.108.264504} {\bibfield  {journal} {\bibinfo  {journal}
  {Physical Review Letters}\ }\textbf {\bibinfo {volume} {108}},\ \bibinfo
  {pages} {264504} (\bibinfo {year} {2012})}\BibitemShut {NoStop}%
\bibitem [{\citenamefont {Bohloli}\ and\ \citenamefont {{de
  Pater}}(2006)}]{bohloli-jpetscieng-2006}%
  \BibitemOpen
  \bibfield  {author} {\bibinfo {author} {\bibfnamefont {B.}~\bibnamefont
  {Bohloli}}\ and\ \bibinfo {author} {\bibfnamefont {C.~J.}\ \bibnamefont {{de
  Pater}}},\ }\bibfield  {title} {\enquote {\bibinfo {title} {Experimental
  study on hydraulic fracturing of soft rocks: {Influence} of fluid rheology
  and confining stress},}\ }\href {\doibase 10.1016/j.petrol.2006.01.009}
  {\bibfield  {journal} {\bibinfo  {journal} {Journal of Petroleum Science and
  Engineering}\ }\textbf {\bibinfo {volume} {53}},\ \bibinfo {pages} {1--12}
  (\bibinfo {year} {2006})}\BibitemShut {NoStop}%
\bibitem [{\citenamefont {Germanovich}\ \emph {et~al.}(2012)\citenamefont
  {Germanovich}, \citenamefont {Hurt}, \citenamefont {Ayoub}, \citenamefont
  {Siebrits}, \citenamefont {Norman}, \citenamefont {Ispas},\ and\
  \citenamefont {Montgomery}}]{germanovich-spe-2012}%
  \BibitemOpen
  \bibfield  {author} {\bibinfo {author} {\bibfnamefont {L.~N.}\ \bibnamefont
  {Germanovich}}, \bibinfo {author} {\bibfnamefont {R.~S.}\ \bibnamefont
  {Hurt}}, \bibinfo {author} {\bibfnamefont {J.~A.}\ \bibnamefont {Ayoub}},
  \bibinfo {author} {\bibfnamefont {E.}~\bibnamefont {Siebrits}}, \bibinfo
  {author} {\bibfnamefont {W.~D.}\ \bibnamefont {Norman}}, \bibinfo {author}
  {\bibfnamefont {I.}~\bibnamefont {Ispas}}, \ and\ \bibinfo {author}
  {\bibfnamefont {C.}~\bibnamefont {Montgomery}},\ }\bibfield  {title}
  {\enquote {\bibinfo {title} {Experimental study of hydraulic fracturing in
  unconsolidated materials},}\ }in\ \href {\doibase 10.2118/151827-MS} {\emph
  {\bibinfo {booktitle} {SPE International Symposium and Exhibition on
  Formation Damage Control, Lafayette, LA, USA (SPE 151827-MS)}}}\ (\bibinfo
  {year} {2012})\BibitemShut {NoStop}%
\bibitem [{\citenamefont {Beavers}\ \emph {et~al.}(1975)\citenamefont
  {Beavers}, \citenamefont {Wilson},\ and\ \citenamefont
  {Masha}}]{beavers-japplmech-1975}%
  \BibitemOpen
  \bibfield  {author} {\bibinfo {author} {\bibfnamefont {G.~S.}\ \bibnamefont
  {Beavers}}, \bibinfo {author} {\bibfnamefont {T.~A.}\ \bibnamefont {Wilson}},
  \ and\ \bibinfo {author} {\bibfnamefont {B.~A.}\ \bibnamefont {Masha}},\
  }\bibfield  {title} {\enquote {\bibinfo {title} {Flow through a deformable
  porous material},}\ }\href {\doibase 10.1115/1.3423648} {\bibfield  {journal}
  {\bibinfo  {journal} {Journal of Applied Mechanics}\ }\textbf {\bibinfo
  {volume} {42}},\ \bibinfo {pages} {598--602} (\bibinfo {year}
  {1975})}\BibitemShut {NoStop}%
\bibitem [{\citenamefont {Parker}\ \emph {et~al.}(1987)\citenamefont {Parker},
  \citenamefont {Mehta},\ and\ \citenamefont {Caro}}]{parker-japplmech-1987}%
  \BibitemOpen
  \bibfield  {author} {\bibinfo {author} {\bibfnamefont {K.~H.}\ \bibnamefont
  {Parker}}, \bibinfo {author} {\bibfnamefont {R.~V.}\ \bibnamefont {Mehta}}, \
  and\ \bibinfo {author} {\bibfnamefont {C.~G.}\ \bibnamefont {Caro}},\
  }\bibfield  {title} {\enquote {\bibinfo {title} {Steady flow in porous,
  elastically deformable materials},}\ }\href {\doibase 10.1115/1.3173119}
  {\bibfield  {journal} {\bibinfo  {journal} {Journal of Applied Mechanics}\
  }\textbf {\bibinfo {volume} {54}},\ \bibinfo {pages} {794} (\bibinfo {year}
  {1987})}\BibitemShut {NoStop}%
\bibitem [{\citenamefont {Sobac}\ \emph {et~al.}(2011)\citenamefont {Sobac},
  \citenamefont {Colombani},\ and\ \citenamefont
  {Forterre}}]{sobac-mecind-2011}%
  \BibitemOpen
  \bibfield  {author} {\bibinfo {author} {\bibfnamefont {B.}~\bibnamefont
  {Sobac}}, \bibinfo {author} {\bibfnamefont {M.}~\bibnamefont {Colombani}}, \
  and\ \bibinfo {author} {\bibfnamefont {Y.}~\bibnamefont {Forterre}},\
  }\bibfield  {title} {\enquote {\bibinfo {title} {On the dynamics of
  poroelastic foams (\emph{in French})},}\ }\href {\doibase
  10.1051/meca/2011115} {\bibfield  {journal} {\bibinfo  {journal}
  {M\'{e}canique \& Industries}\ }\textbf {\bibinfo {volume} {12}},\ \bibinfo
  {pages} {231--238} (\bibinfo {year} {2011})}\BibitemShut {NoStop}%
\bibitem [{\citenamefont {Mukhopadhyay}\ and\ \citenamefont
  {Peixinho}(2011)}]{mukhopadhyay-pre-2011}%
  \BibitemOpen
  \bibfield  {author} {\bibinfo {author} {\bibfnamefont {S.}~\bibnamefont
  {Mukhopadhyay}}\ and\ \bibinfo {author} {\bibfnamefont {J.}~\bibnamefont
  {Peixinho}},\ }\bibfield  {title} {\enquote {\bibinfo {title} {Packings of
  deformable spheres},}\ }\href {\doibase
  http://link.aps.org/doi/10.1103/PhysRevE.84.011302} {\bibfield  {journal}
  {\bibinfo  {journal} {Physical Review E}\ }\textbf {\bibinfo {volume} {84}},\
  \bibinfo {pages} {011302} (\bibinfo {year} {2011})}\BibitemShut {NoStop}%
\bibitem [{\citenamefont {Brodu}\ \emph {et~al.}()\citenamefont {Brodu},
  \citenamefont {Dijksman},\ and\ \citenamefont
  {Behringer}}]{brodu-arxiv-2014}%
  \BibitemOpen
  \bibfield  {author} {\bibinfo {author} {\bibfnamefont {N.}~\bibnamefont
  {Brodu}}, \bibinfo {author} {\bibfnamefont {J.~A.}\ \bibnamefont {Dijksman}},
  \ and\ \bibinfo {author} {\bibfnamefont {R.~P.}\ \bibnamefont {Behringer}},\
  }\href@noop {} {\enquote {\bibinfo {title} {Spanning the scales of granular
  materials: {Microscopic} force imaging},}\ }\bibinfo {note} {Available at
  {http://arxiv.org/abs/1408.2506}}\BibitemShut {NoStop}%
\bibitem [{\citenamefont {Zhang}\ and\ \citenamefont
  {Huang}(2011)}]{zhang-arma-2011}%
  \BibitemOpen
  \bibfield  {author} {\bibinfo {author} {\bibfnamefont {F.}~\bibnamefont
  {Zhang}}\ and\ \bibinfo {author} {\bibfnamefont {H.}~\bibnamefont {Huang}},\
  }\bibfield  {title} {\enquote {\bibinfo {title} {Coupled {DEM}-{CFD} modeling
  of fluid injection into granular media},}\ }in\ \href {\doibase
  https://www.onepetro.org/conference-paper/ARMA-11-228} {\emph {\bibinfo
  {booktitle} {45th US Rock Mechanics / Geomechanics Symposium, San Francisco,
  CA, USA}}}\ (\bibinfo {year} {2011})\ \bibinfo {note} {ARMA
  11-228}\BibitemShut {NoStop}%
\bibitem [{\citenamefont {Pinto}\ \emph {et~al.}(2007)\citenamefont {Pinto},
  \citenamefont {Couto}, \citenamefont {Atman}, \citenamefont {Alves},
  \citenamefont {Bernardes}, \citenamefont {{de Resende}},\ and\ \citenamefont
  {Souza}}]{pinto-prl-2007}%
  \BibitemOpen
  \bibfield  {author} {\bibinfo {author} {\bibfnamefont {S.~F.}\ \bibnamefont
  {Pinto}}, \bibinfo {author} {\bibfnamefont {M.~S.}\ \bibnamefont {Couto}},
  \bibinfo {author} {\bibfnamefont {A.~P.~F.}\ \bibnamefont {Atman}}, \bibinfo
  {author} {\bibfnamefont {S.~G.}\ \bibnamefont {Alves}}, \bibinfo {author}
  {\bibfnamefont {A.~T.}\ \bibnamefont {Bernardes}}, \bibinfo {author}
  {\bibfnamefont {H.~F.~V.}\ \bibnamefont {{de Resende}}}, \ and\ \bibinfo
  {author} {\bibfnamefont {E.~C.}\ \bibnamefont {Souza}},\ }\bibfield  {title}
  {\enquote {\bibinfo {title} {Granular fingers on jammed systems: {N}ew
  fluidlike patterns arising in grain-grain invasion experiments},}\ }\href
  {\doibase 10.1103/PhysRevLett.99.068001} {\bibfield  {journal} {\bibinfo
  {journal} {Physical Review Letters}\ }\textbf {\bibinfo {volume} {99}},\
  \bibinfo {pages} {068001} (\bibinfo {year} {2007})}\BibitemShut {NoStop}%
\bibitem [{\citenamefont {Falk}\ and\ \citenamefont
  {Langer}(1998)}]{falk-pre-1998}%
  \BibitemOpen
  \bibfield  {author} {\bibinfo {author} {\bibfnamefont {M.~L.}\ \bibnamefont
  {Falk}}\ and\ \bibinfo {author} {\bibfnamefont {J.~S.}\ \bibnamefont
  {Langer}},\ }\bibfield  {title} {\enquote {\bibinfo {title} {Dynamics of
  viscoplastic deformation in amorphous solids},}\ }\href {\doibase
  10.1103/PhysRevE.57.7192} {\bibfield  {journal} {\bibinfo  {journal}
  {Physical Review E}\ }\textbf {\bibinfo {volume} {57}},\ \bibinfo {pages}
  {7192--7205} (\bibinfo {year} {1998})}\BibitemShut {NoStop}%
\bibitem [{\citenamefont {Lundberg}\ \emph {et~al.}(2008)\citenamefont
  {Lundberg}, \citenamefont {Krishan}, \citenamefont {Xu}, \citenamefont
  {{O'Hern}},\ and\ \citenamefont {Dennin}}]{lundberg-pre-2008}%
  \BibitemOpen
  \bibfield  {author} {\bibinfo {author} {\bibfnamefont {M.}~\bibnamefont
  {Lundberg}}, \bibinfo {author} {\bibfnamefont {K.}~\bibnamefont {Krishan}},
  \bibinfo {author} {\bibfnamefont {N.}~\bibnamefont {Xu}}, \bibinfo {author}
  {\bibfnamefont {C.~S.}\ \bibnamefont {{O'Hern}}}, \ and\ \bibinfo {author}
  {\bibfnamefont {M.}~\bibnamefont {Dennin}},\ }\bibfield  {title} {\enquote
  {\bibinfo {title} {Reversible plastic events in amorphous materials},}\
  }\href {\doibase 10.1103/PhysRevE.77.041505} {\bibfield  {journal} {\bibinfo
  {journal} {Physical Review E}\ }\textbf {\bibinfo {volume} {77}},\ \bibinfo
  {pages} {041505} (\bibinfo {year} {2008})}\BibitemShut {NoStop}%
\bibitem [{\citenamefont {Keim}\ and\ \citenamefont
  {Arratia}(2013)}]{keim-softmatter-2013}%
  \BibitemOpen
  \bibfield  {author} {\bibinfo {author} {\bibfnamefont {N.~C.}\ \bibnamefont
  {Keim}}\ and\ \bibinfo {author} {\bibfnamefont {P.~E.}\ \bibnamefont
  {Arratia}},\ }\bibfield  {title} {\enquote {\bibinfo {title} {Yielding and
  microstructure in a {2D} jammed material under shear deformation},}\ }\href
  {\doibase 10.1039/C3SM51014J} {\bibfield  {journal} {\bibinfo  {journal}
  {Soft Matter}\ }\textbf {\bibinfo {volume} {9}},\ \bibinfo {pages}
  {6222--6225} (\bibinfo {year} {2013})}\BibitemShut {NoStop}%
\bibitem [{\citenamefont {Keim}\ and\ \citenamefont
  {Arratia}(2014)}]{keim-prl-2014}%
  \BibitemOpen
  \bibfield  {author} {\bibinfo {author} {\bibfnamefont {N.~C.}\ \bibnamefont
  {Keim}}\ and\ \bibinfo {author} {\bibfnamefont {P.~E.}\ \bibnamefont
  {Arratia}},\ }\bibfield  {title} {\enquote {\bibinfo {title} {Mechanical and
  microscopic properties of the reversible plastic regime in a {2D} jammed
  material},}\ }\href {\doibase 10.1103/PhysRevLett.112.028302} {\bibfield
  {journal} {\bibinfo  {journal} {Physical Review Letters}\ }\textbf {\bibinfo
  {volume} {112}},\ \bibinfo {pages} {028302} (\bibinfo {year}
  {2014})}\BibitemShut {NoStop}%
\bibitem [{\citenamefont {Li\'{e}tor-Santos}\ \emph {et~al.}(2011)\citenamefont
  {Li\'{e}tor-Santos}, \citenamefont {Sierra-Mart\'{i}n},\ and\ \citenamefont
  {Fern\'{a}ndez-Nieves}}]{lietor-santos-pre-2011}%
  \BibitemOpen
  \bibfield  {author} {\bibinfo {author} {\bibfnamefont {J.~J.}\ \bibnamefont
  {Li\'{e}tor-Santos}}, \bibinfo {author} {\bibfnamefont {B.}~\bibnamefont
  {Sierra-Mart\'{i}n}}, \ and\ \bibinfo {author} {\bibfnamefont
  {A.}~\bibnamefont {Fern\'{a}ndez-Nieves}},\ }\bibfield  {title} {\enquote
  {\bibinfo {title} {Bulk and shear moduli of compressed microgel
  suspensions},}\ }\href {\doibase 10.1103/PhysRevE.84.060402} {\bibfield
  {journal} {\bibinfo  {journal} {Physical Review E}\ }\textbf {\bibinfo
  {volume} {84}},\ \bibinfo {pages} {060402(R)} (\bibinfo {year} {2011})}\BibitemShut {NoStop}%
\bibitem [{\citenamefont {O'Hern}\ \emph {et~al.}(2003)\citenamefont {O'Hern},
  \citenamefont {Silbert}, \citenamefont {Liu},\ and\ \citenamefont
  {Nagel}}]{ohern-pre-2003}%
  \BibitemOpen
  \bibfield  {author} {\bibinfo {author} {\bibfnamefont {C.~S.}\ \bibnamefont
  {O'Hern}}, \bibinfo {author} {\bibfnamefont {L.~E.}\ \bibnamefont {Silbert}},
  \bibinfo {author} {\bibfnamefont {A.~J.}\ \bibnamefont {Liu}}, \ and\
  \bibinfo {author} {\bibfnamefont {S.~R.}\ \bibnamefont {Nagel}},\ }\bibfield
  {title} {\enquote {\bibinfo {title} {Jamming at zero temperature and zero
  applied stress: {T}he epitome of disorder},}\ }\href {\doibase
  10.1103/PhysRevE.68.011306} {\bibfield  {journal} {\bibinfo  {journal}
  {Physical Review E}\ }\textbf {\bibinfo {volume} {68}},\ \bibinfo {pages}
  {011306} (\bibinfo {year} {2003})}\BibitemShut {NoStop}%
\bibitem [{\citenamefont {Nordstrom}\ \emph {et~al.}(2010)\citenamefont
  {Nordstrom}, \citenamefont {Verneuil}, \citenamefont {Arratia}, \citenamefont
  {Basu}, \citenamefont {Zhang}, \citenamefont {Yodh}, \citenamefont {Gollub},\
  and\ \citenamefont {Durian}}]{nordstrom-prl-2010}%
  \BibitemOpen
  \bibfield  {author} {\bibinfo {author} {\bibfnamefont {K.~N.}\ \bibnamefont
  {Nordstrom}}, \bibinfo {author} {\bibfnamefont {E.}~\bibnamefont {Verneuil}},
  \bibinfo {author} {\bibfnamefont {P.~E.}\ \bibnamefont {Arratia}}, \bibinfo
  {author} {\bibfnamefont {A.}~\bibnamefont {Basu}}, \bibinfo {author}
  {\bibfnamefont {Z.}~\bibnamefont {Zhang}}, \bibinfo {author} {\bibfnamefont
  {A.~G.}\ \bibnamefont {Yodh}}, \bibinfo {author} {\bibfnamefont {J.~P.}\
  \bibnamefont {Gollub}}, \ and\ \bibinfo {author} {\bibfnamefont {D.~J.}\
  \bibnamefont {Durian}},\ }\bibfield  {title} {\enquote {\bibinfo {title}
  {Microfluidic rheology of soft colloids above and below {J}amming},}\ }\href
  {\doibase 10.1103/PhysRevLett.105.175701} {\bibfield  {journal} {\bibinfo
  {journal} {Physical Review Letters}\ }\textbf {\bibinfo {volume} {105}},\
  \bibinfo {pages} {175701} (\bibinfo {year} {2010})}\BibitemShut {NoStop}%
\bibitem [{\citenamefont {Boyer}\ \emph {et~al.}(2011)\citenamefont {Boyer},
  \citenamefont {Guazzelli},\ and\ \citenamefont {Pouliquen}}]{boyer-prl-2011}%
  \BibitemOpen
  \bibfield  {author} {\bibinfo {author} {\bibfnamefont {F.}~\bibnamefont
  {Boyer}}, \bibinfo {author} {\bibfnamefont {\'{E}.}\ \bibnamefont
  {Guazzelli}}, \ and\ \bibinfo {author} {\bibfnamefont {O.}~\bibnamefont
  {Pouliquen}},\ }\bibfield  {title} {\enquote {\bibinfo {title} {Unifying
  suspension and granular rheology},}\ }\href {\doibase
  10.1103/PhysRevLett.107.188301} {\bibfield  {journal} {\bibinfo  {journal}
  {Physical Review Letters}\ }\textbf {\bibinfo {volume} {107}},\ \bibinfo
  {pages} {188301} (\bibinfo {year} {2011})}\BibitemShut {NoStop}%
\bibitem [{\citenamefont {Bandi}\ \emph {et~al.}(2011)\citenamefont {Bandi},
  \citenamefont {Tallinen},\ and\ \citenamefont {Mahadevan}}]{bandi-epl-2011}%
  \BibitemOpen
  \bibfield  {author} {\bibinfo {author} {\bibfnamefont {M.~M.}\ \bibnamefont
  {Bandi}}, \bibinfo {author} {\bibfnamefont {T.}~\bibnamefont {Tallinen}}, \
  and\ \bibinfo {author} {\bibfnamefont {L.}~\bibnamefont {Mahadevan}},\
  }\bibfield  {title} {\enquote {\bibinfo {title} {Shock-driven jamming and
  periodic fracture of particulate rafts},}\ }\href {\doibase
  10.1209/0295-5075/96/36008} {\bibfield  {journal} {\bibinfo  {journal} {EPL
  (Europhysics Letters)}\ }\textbf {\bibinfo {volume} {96}},\ \bibinfo {pages}
  {36008} (\bibinfo {year} {2011})}\BibitemShut {NoStop}%
\bibitem [{\citenamefont {Barry}\ and\ \citenamefont
  {Aldis}(1993)}]{barry-jaustralmathsocb-1993}%
  \BibitemOpen
  \bibfield  {author} {\bibinfo {author} {\bibfnamefont {S.~I.}\ \bibnamefont
  {Barry}}\ and\ \bibinfo {author} {\bibfnamefont {G.~K.}\ \bibnamefont
  {Aldis}},\ }\bibfield  {title} {\enquote {\bibinfo {title} {Radial flow
  through deformable porous shells},}\ }\href {\doibase
  10.1017/S0334270000008936} {\bibfield  {journal} {\bibinfo  {journal}
  {Journal of the Australian Mathematical Society. Series B. Applied
  Mathematics}\ }\textbf {\bibinfo {volume} {34}},\ \bibinfo {pages} {333--354}
  (\bibinfo {year} {1993})}\BibitemShut {NoStop}%
\bibitem [{\citenamefont {Falk}\ and\ \citenamefont
  {Langer}(2011)}]{falk-annrevcondmattphys-2011}%
  \BibitemOpen
  \bibfield  {author} {\bibinfo {author} {\bibfnamefont {M.~L.}\ \bibnamefont
  {Falk}}\ and\ \bibinfo {author} {\bibfnamefont {J.~S.}\ \bibnamefont
  {Langer}},\ }\bibfield  {title} {\enquote {\bibinfo {title} {Deformation and
  failure of amorphous, solidlike materials},}\ }\href {\doibase
  10.1146/annurev-conmatphys-062910-140452} {\bibfield  {journal} {\bibinfo
  {journal} {Annual Review of Condensed Matter Physics}\ }\textbf {\bibinfo
  {volume} {2}},\ \bibinfo {pages} {353--373} (\bibinfo {year}
  {2011})}\BibitemShut {NoStop}%
\bibitem [{\citenamefont {Dijksman}\ \emph {et~al.}(2012)\citenamefont
  {Dijksman}, \citenamefont {Rietz}, \citenamefont {L\"{o}rincz}, \citenamefont
  {{van Hecke}},\ and\ \citenamefont {Losert}}]{dijksman-rsi-2012}%
  \BibitemOpen
  \bibfield  {author} {\bibinfo {author} {\bibfnamefont {J.~A.}\ \bibnamefont
  {Dijksman}}, \bibinfo {author} {\bibfnamefont {F.}~\bibnamefont {Rietz}},
  \bibinfo {author} {\bibfnamefont {K.~A.}\ \bibnamefont {L\"{o}rincz}},
  \bibinfo {author} {\bibfnamefont {M.}~\bibnamefont {{van Hecke}}}, \ and\
  \bibinfo {author} {\bibfnamefont {W.}~\bibnamefont {Losert}},\ }\bibfield
  {title} {\enquote {\bibinfo {title} {Invited {Article}: {R}efractive index
  matched scanning of dense granular materials},}\ }\href {\doibase
  10.1063/1.3674173} {\bibfield  {journal} {\bibinfo  {journal} {Review of
  Scientific Instruments}\ }\textbf {\bibinfo {volume} {83}},\ \bibinfo {pages}
  {011301} (\bibinfo {year} {2012})}\BibitemShut {NoStop}%
\bibitem [{\citenamefont {Byron}\ and\ \citenamefont
  {Variano}(2013)}]{byron-expfluids-2013}%
  \BibitemOpen
  \bibfield  {author} {\bibinfo {author} {\bibfnamefont {M.~L.}\ \bibnamefont
  {Byron}}\ and\ \bibinfo {author} {\bibfnamefont {E.~A.}\ \bibnamefont
  {Variano}},\ }\bibfield  {title} {\enquote {\bibinfo {title}
  {Refractive-index-matched hydrogel materials for measuring flow-structure
  interactions},}\ }\href {\doibase 10.1007/s00348-013-1456-z} {\bibfield
  {journal} {\bibinfo  {journal} {Experiments in Fluids}\ }\textbf {\bibinfo
  {volume} {54}},\ \bibinfo {pages} {1456} (\bibinfo {year}
  {2013})}\BibitemShut {NoStop}%
\bibitem [{\citenamefont {Bragg}\ and\ \citenamefont
  {Nye}(1947)}]{bragg-procrsoca-1947}%
  \BibitemOpen
  \bibfield  {author} {\bibinfo {author} {\bibfnamefont {L.}~\bibnamefont
  {Bragg}}\ and\ \bibinfo {author} {\bibfnamefont {J.~F.}\ \bibnamefont
  {Nye}},\ }\bibfield  {title} {\enquote {\bibinfo {title} {A dynamical model
  of a crystal structure},}\ }\href {\doibase 10.1098/rspa.1947.0089}
  {\bibfield  {journal} {\bibinfo  {journal} {Proceedings of the Royal Society
  A}\ }\textbf {\bibinfo {volume} {190}},\ \bibinfo {pages} {474--481}
  (\bibinfo {year} {1947})}\BibitemShut {NoStop}%
\bibitem [{\citenamefont {Argon}\ and\ \citenamefont
  {Kuo}(1979)}]{argon-matscieng-1979}%
  \BibitemOpen
  \bibfield  {author} {\bibinfo {author} {\bibfnamefont {A.~S.}\ \bibnamefont
  {Argon}}\ and\ \bibinfo {author} {\bibfnamefont {H.~Y.}\ \bibnamefont
  {Kuo}},\ }\bibfield  {title} {\enquote {\bibinfo {title} {Plastic flow in a
  disordered bubble raft (an analog of a metallic glass)},}\ }\href {\doibase
  10.1016/0025-5416(79)90174-5} {\bibfield  {journal} {\bibinfo  {journal}
  {Materials Science and Engineering}\ }\textbf {\bibinfo {volume} {39}},\
  \bibinfo {pages} {101--109} (\bibinfo {year} {1979})}\BibitemShut {NoStop}%
\bibitem [{\citenamefont {Skjeltorp}\ and\ \citenamefont
  {Meakin}(1988)}]{skjeltorp-nature-1988}%
  \BibitemOpen
  \bibfield  {author} {\bibinfo {author} {\bibfnamefont {A.~T.}\ \bibnamefont
  {Skjeltorp}}\ and\ \bibinfo {author} {\bibfnamefont {P.}~\bibnamefont
  {Meakin}},\ }\bibfield  {title} {\enquote {\bibinfo {title} {Fracture in
  microsphere monolayers studied by experiment and computer simulation},}\
  }\href {\doibase 10.1038/335424a0} {\bibfield  {journal} {\bibinfo  {journal}
  {Nature}\ }\textbf {\bibinfo {volume} {335}},\ \bibinfo {pages} {424--426}
  (\bibinfo {year} {1988})}\BibitemShut {NoStop}%
\bibitem [{\citenamefont {Meakin}(1991)}]{meakin-science-1991}%
  \BibitemOpen
  \bibfield  {author} {\bibinfo {author} {\bibfnamefont {P.}~\bibnamefont
  {Meakin}},\ }\bibfield  {title} {\enquote {\bibinfo {title} {Models for
  material failure and deformation},}\ }\href {\doibase
  10.1126/science.252.5003.226} {\bibfield  {journal} {\bibinfo  {journal}
  {Science}\ }\textbf {\bibinfo {volume} {252}},\ \bibinfo {pages} {226--234}
  (\bibinfo {year} {1991})}\BibitemShut {NoStop}%
\bibitem [{\citenamefont {Holtzman}\ and\ \citenamefont
  {Juanes}(2010)}]{holtzman-pre-2010}%
  \BibitemOpen
  \bibfield  {author} {\bibinfo {author} {\bibfnamefont {R.}~\bibnamefont
  {Holtzman}}\ and\ \bibinfo {author} {\bibfnamefont {R.}~\bibnamefont
  {Juanes}},\ }\bibfield  {title} {\enquote {\bibinfo {title} {Crossover from
  fingering to fracturing in deformable disordered media},}\ }\href {\doibase
  10.1103/PhysRevE.82.046305} {\bibfield  {journal} {\bibinfo  {journal}
  {Physical Review E}\ }\textbf {\bibinfo {volume} {82}},\ \bibinfo {pages}
  {046305} (\bibinfo {year} {2010})}\BibitemShut {NoStop}%
\bibitem [{\citenamefont {Zhang}\ \emph {et~al.}(2013)\citenamefont {Zhang},
  \citenamefont {Damjanac},\ and\ \citenamefont {Huang}}]{zhang-jgr-2013}%
  \BibitemOpen
  \bibfield  {author} {\bibinfo {author} {\bibfnamefont {F.}~\bibnamefont
  {Zhang}}, \bibinfo {author} {\bibfnamefont {B.}~\bibnamefont {Damjanac}}, \
  and\ \bibinfo {author} {\bibfnamefont {H.}~\bibnamefont {Huang}},\ }\bibfield
   {title} {\enquote {\bibinfo {title} {Coupled discrete element modeling of
  fluid injection into dense granular media},}\ }\href {\doibase
  10.1002/jgrb.50204} {\bibfield  {journal} {\bibinfo  {journal} {Journal of
  Geophysical Research}\ }\textbf {\bibinfo {volume} {118}},\ \bibinfo {pages}
  {2703--2722} (\bibinfo {year} {2013})}\BibitemShut {NoStop}%
\bibitem [{\citenamefont {Blair}\ and\ \citenamefont
  {Dufresne}()}]{blair-code}%
  \BibitemOpen
  \bibfield  {author} {\bibinfo {author} {\bibfnamefont {D.~L.}\ \bibnamefont
  {Blair}}\ and\ \bibinfo {author} {\bibfnamefont {E.~R.}\ \bibnamefont
  {Dufresne}},\ }\href@noop {} {\enquote {\bibinfo {title} {The matlab particle
  tracking code repository},}\ }\bibinfo {note} {Available at
  {http://physics.georgetown.edu/matlab/}}\BibitemShut {NoStop}%
\end{thebibliography}

%

\end{document}